\documentclass[11pt]{article}
   
\usepackage[margin=.75in]{geometry}

\usepackage{amsmath,amssymb}
\usepackage{cancel}
\usepackage{graphicx}
\usepackage{caption}
\usepackage{comment}
\usepackage{subcaption}

\usepackage{nicematrix}
\usepackage{bm}

\usepackage{color}

\definecolor{indigo}{RGB}{0,0,120}
\usepackage[colorlinks=true, linkcolor=indigo, citecolor=blue, urlcolor=indigo,backref=page]{hyperref}

\def\tr{\,{\rm tr}\,}

\def\fl{\noindent}

\newcommand{\bra}{\langle}
\newcommand{\ket}{\rangle}

\newcommand{\tl}[1]{\tilde{#1}}

\newcommand{\DD}[2]{\frac {d #1}{d #2}}

\newcommand{\beq}{\begin{equation}}
\newcommand{\eeq}{\end{equation}}
\newcommand{\beqs}{\begin{eqnarray}}
\newcommand{\eeqs}{\end{eqnarray}}
\newcommand{\half}{\frac{1}{2}}
\newcommand{\ov}[1]{\frac{1}{#1}}

\def\al{\alpha} 		
\def\del{\delta}
\def\D{\Delta}	
 
\def\eps{\epsilon} 
\def\la{\lambda}
		
\def\sig{\sigma}
\def\tht{\theta}
\def\om{\omega}

\newcount\colveccount  
\newcommand*\colvec[1]{\global\colveccount#1  \begin{pmatrix} \colvecnext} \def\colvecnext#1{#1 \global\advance\colveccount-1
        \ifnum\colveccount>0 \\ \expandafter\colvecnext
        \else \end{pmatrix} \fi}

\usepackage{calligra} 
\DeclareMathAlphabet{\mathcalligra}{T1}{calligra}{m}{n}
\DeclareFontShape{T1}{calligra}{m}{n}{<->s*[2.2]callig15}{}

\begin{document}

\title{
\hfill{\tt \small \href{https://arxiv.org/abs/2312.13122}{arXiv:2312.13122[nlin.SI]}}\\
Screwon spectral statistics and dispersion relation in the quantum Rajeev-Ranken model} 
\author{{\sc Govind S. Krishnaswami$^{a}$ and T. R. Vishnu$^{a, b}$}
\\ \small
$^{a}$Physics Department, Chennai Mathematical Institute,  SIPCOT IT Park, Siruseri 603103, India\\ \small
$^{b}$Physics Department, Raman Research Institute,  Sadashivanagar, Bengaluru 560080, India\\
\\ \small
Email: {\tt govind@cmi.ac.in, vishnu@rrimail.rri.res.in}}
\date{April 21, 2024\\ 
\vspace{0.1cm}
Published in \href{https://doi.org/10.1016/j.physd.2024.134170}{Physica D 463, 134170 (2024)}}

\maketitle

\vspace{-1cm}

\abstract{ \small The Rajeev-Ranken (RR) model is a Hamiltonian system describing screw-type nonlinear waves (screwons) of wavenumber $k$ in a scalar field theory pseudodual to the 1+1D SU(2) principal chiral model. Classically, the RR model based on a quadratic Hamiltonian on a nilpotent/Euclidean Poisson algebra is Liouville integrable. Upon adopting canonical variables in a slightly extended phase space, the model was interpreted as a novel 3D cylindrically symmetric quartic oscillator with a rotational energy. Here, we examine the spectral statistics and dispersion relation of quantized screwons via numerical diagonalization validated by variational and perturbative approximations. We also derive a semiclassical estimate for the cumulative level distribution which compares favorably with the one from numerical diagonalization. The spectrum shows level crossings typical of an integrable system. The $i^{\rm th}$ unfolded nearest neighbor spacings are found to follow Poisson statistics for small $i$. Nonoverlapping spacing ratios also indicate that successive spectral gaps are independently distributed. After displaying universal linear behavior over energy windows of short lengths, the spectral rigidity saturates at a length and value that scales with the square-root of energy. For strong coupling $\lambda$ and intermediate $k$, we argue that reduced screwon energies can depend only on the product $\lambda k$. Numerically, we find power law dependences on $\lambda$ and $k$ with an approximately common exponent $2/3$ provided the angular momentum quantum number $l$ is small compared to the number of nodes $n$ in the radial wavefunction. On the other hand, for the ground state $n = l = 0$, the common exponent becomes 1.}

\vspace{0.25 cm}
\footnotesize


\normalsize

{\scriptsize \tableofcontents}

\normalsize

\section{Introduction}
\label{s:Introduction}

The Rajeev-Ranken (RR) model \cite{R-R} is a mechanical system with three degrees of freedom. It describes nonlinear screw-type continuous waves  in a $1+1$-dimensional nilpotent scalar field theory which is dual to the SU(2)-principal chiral model \cite{Polyakov}. The scalar field $\phi$ is valued in the $\mathfrak{su}$(2) Lie algebra and classically satisfies the quadratically nonlinear evolution equations 
	\beq
	\ddot{\phi} = \phi'' + \la [\dot{\phi}, \phi']
	\label{e:EOM-scalar-field-theory}
	\eeq
	for a dimensionless coupling constant $\la$. The screw-type waves are solutions of the form 
	\beq
	\phi(x,t) = e^{Kx}R(t)e^{-Kx} + m K x,
	\label{e:screwon-ansatz}
	\eeq
where $R(t)$ is a dynamical $\mathfrak{su}$(2) matrix  with the three real degrees of freedom $R_a = i\tr(R \sig_a)$ for $a=1,2,3$. Moreover, $K = i k \sig_3/2$ with $k$ a constant with dimensions of a wavenumber and $m$ is a dimensionless real parameter. The resulting equations of motion for $R$ are elegantly expressed as
	\beq
	\dot L = [K, S] \quad \text{and} \quad \dot S = \la [S, L] \quad \text{where} \quad
	L = [K, R] + m K \quad \text{and} \quad S = \dot{R} + \frac{K}{\la}.
	\eeq
These equations follow from a quadratic Hamiltonian and a 	nilpotent Poisson algebra among $L$ and $S$ in a six dimensional phase space with coordinates $L_a = i\tr(L \sig_a)$ and $S_a = i\tr(S \sig_a)$. They also admit a Lax pair formulation and classical $r$-matrix leading to a complete set of independent conserved quantities in involution implying Liouville integrability \cite{G-V-1, G-V-2}. However, $L_3 = -m k$ is a Casimir invariant of the nilpotent algebra and $R_3$ does not appear in the $L-S$ equations of motion. The evolution of $R_3$ is determined by $\dot{R_3} = S_3 + k/\la$. Thus in this paper, we will treat $L_3$ as a parameter  and introduce $R_3$ as a coordinate. The equations of motion of the RR model admit an alternate Hamitonian-Poisson bracket formulation in terms of canonical variables $R_{1,2,3}$ and their conjugate momenta \cite{G-V-1, G-V-2}:
	\beq
	kP_{1,2} = \dot R_{1,2} \pm \half \la k m R_{2,1} \quad \text{and} \quad kP_3 = \dot R_3 + \half \la k (R_1^2 + R_2^2).
	\label{e:conjugate-momenta}
	\eeq 
The Hamiltonian 
	\beqs
	 H &=& \half \left[ \left( k P_1 - \frac{\la k m R_2}{2} \right)^2 + \left( k P_2 + \frac{\la k m R_1}{2} \right)^2 + \left( k P_3 - \frac{\la k}{2}(R_1^2 + R_2^2) \right)^2 \right] \cr
	 && + \frac{k^2}{2}(R_1^2 + R_2^2 + m^2),
	\label{e:Hamiltonian-mech-RP}
	\eeqs
may be interpreted as that of a cylindrically symmetric anharmonic oscillator in three dimensions \cite{G-V-3}. Indeed, if we regard $R_{1, 2, 3}$ and $k P_{1,2,3}$ as the Cartesian components of the position and momentum of a particle of mass $\mu = 1$, then (\ref{e:Hamiltonian-mech-RP}) is equivalent to the Hamiltonian  
	\beqs
	H &=& \frac{p_x^2 + p_y^2 + p_z^2}{2\mu} + \frac{\la k m (x p_y - y p_x)}{2\mu} + \left( \frac{\la^2 k^2 m^2 }{8 \mu} - \frac{\la k p_z}{2\mu} + \frac{k^2}{2} \right) (x^2 + y^2) \cr
	&& + \frac{\la^2 k^2}{8 \mu} (x^2 + y^2)^2 + \frac{k^2 m^2}{2}.
	\label{e:Hamiltonian-quadratic-quartic-Cartesian}
	\eeqs
This Hamiltonian describes a particle of mass $\mu$ moving in a cylindrically symmetric quadratic plus quartic potential with an additional rotational energy proportional to $L_z = x p_y - y p_x$. 

Upon canonical quantization, the cylindrical symmetry of $H$ permits separation of variables in the Schr\"odinger equation for the wavefuntion
	\beq
	\psi(r, \tht, z) = \frac{1}{\sqrt{r}} \varrho(r) \exp{(i l \tht)} \exp{\left(\frac{i p_z z}{\hbar}\right)} \quad \text{for} \quad r \geq 0.
	\label{e:product-wavefunction}
	\eeq
Here, $l \hbar$ is the eigenvalue of $L_z$ and can be any integer while $p_z \in \mathbb{R}$ is the eigenvalue of the operator $p_z$, both of which commute with the Hamiltonian. Division by $\sqrt{r}$ eliminates the $\varrho'$ term in the resulting radial eigenvalue equation
	\beq
	-\frac{\hbar^2}{2 \mu} \varrho''(r) + \left(  U(r) + \frac{\hbar^2}{2 \mu r^2} \left( l^2 - \frac{1}{4}\right)  + \frac{\hbar l \la k m}{2 \mu} + \frac{p_z^2}{2 \mu} + \frac{k^2 m^2}{2} \right) \varrho = E \varrho.
	\label{e:radial-equation-dimensionful-varrho}
	\eeq
Here, 
	\beq
	U(r) = \al r^2 + \beta r^4 \quad \text{where} \quad 
	\al =  \frac{\la^2 k^2 m^2}{8 \mu} - \frac{\la k p_z}{2\mu} + \frac{k^2}{2} \quad \text{and} \quad \beta= \frac{\la^2 k^2}{8 \mu}.
	\label{e:alpha-and-beta}
	\eeq
The parameters $m$ and $k^2$ have dimensions of length and energy/area, while $\la^2$ has dimensions of mass/area. Although $k$ does not have dimensions of inverse length, we will refer to it as a wavenumber since it arose as a wavenumber in the screwon ansatz (\ref{e:screwon-ansatz}). In what follows, we will often work in units of mass, length and angular momentum where $\mu =1$, $m=1$ and $\hbar =1$. The Hamiltonian in Eqn.(\ref{e:Hamiltonian-quadratic-quartic-Cartesian}) is the sum of squares of Hermitian operators. Thus, the energy eigenvalues must satisfy $E \geq 0$. These are the energies of the quantized screw-type waves which we refer to as screwons. Due to translational invariance in the $z$-direction, $E$ includes a free particle contribution $p_z^2/ 2 \mu$. We will mostly be interested in the contribution to the energies coming from the dynamics in the $x$-$y$ plane which for fixed $p_z$ is discrete and labelled by $l$ and a radial quantum number $n$. Aside from $p_z$ these energy levels $E_{n,l}$ depend parametrically on the coupling $\la$ and the screwon wavenumber $k$. Unlike quantized small oscillations around the vacuum of the scalar field theory, screwons are new nonlinear degrees of freedom that could play a role similar to solitons in other field theories. Thus it is interesting to understand the spectrum of screwons, their shapes, dispersion relations (dependence on $k$) and their behavior at strong coupling $\la$, which corresponds to high energies in the field theory. 

In this article, we study the spectral statistics and dispersion relations of quantized screwons. In Section \ref{s:variational-principle}, we use the (radial) asymptotic behaviour known from \cite{G-V-3} to propose an ansatz for the screwon ground state wavefunction and obtain a variational estimate for the ground state energy. In Section \ref{s:First-order-perturbation-theory}, we use first order perturbation theory to determine the dependences of $E_{0, l}$ on $l$ and $\la$ at weak coupling. In Section \ref{s:finite-difference method}, we find the screwon spectrum by numerical diagonalization of the radial Hamiltonian for each fixed $l$. While we do this for a range of values of $k$ and $\la$, we only consider the case $p_z =1$. In fact, for large coupling and an appropriate range of moderate $k$, we argue that the shifted spectrum ($E - p_z^2/2\mu - m^2 k^2/2$) is independent $p_z$. Combining the contributing $l$ sectors we determine the lowest $2\times 10^4$ energy levels. To do this accurately, we optimize the spatial grid (for each value of $k$ and $\la$) to capture the tail as well as all oscillations of the radial wavefunction. The numerical results are validated using our variational estimate and perturbative estimate at weak coupling. As expected of a classically integrable system, the screwon energies display level crossings as $\la$ is varied.

Section \ref{s:energy-level-statistics} is devoted to the level statistics of the screwon spectrum. We express the cumulative level distribution $n(E)$ as a sum of contributions $n(E; p_{\tht})$ from all allowed angular momentum sectors. Semiclassical estimates for $n(E; p_{\tht})$ and $n(E)$ match those obtained from numerical diagonalization for all but the lowest energies. Asymptotically, we find that $n(E)$ displays power law growth while $n(E; p_\tht) \sim (E - E_{0,l})^{\zeta(|p_{\tht}|)}$, with $\zeta$ increasing monotonically with $|p_{\tht}|$. Here $E_{0,l}$ is the ground state energy for $p_{\tht} =l \hbar$. We then examine spacing distributions of the screwon spectrum. The $i^{\rm th}$ unfolded nearest neighbour spacing distributions follow Poisson statistics although deviations become more pronounced as $i$ increases. The independence in the distribution of spectral gaps is also confirmed using the statistics of nonoverlapping spacing ratios. Both these are as expected of a classically integrable system. Number variance $\Sigma(E, L)$ and spectral rigidity $\D(E, L)$, which encode fluctuations in the spectrum around a central energy $E$, display universal linear behaviour over energy windows of short lengths $L$. For larger $L$, they saturate and display oscillations that should reveal system specific details.  Both the value of $\D$ and $L$ at which the saturation occurs show square-root power law behavior as a function of central energy $E$. 

In Section \ref{s:dispersion-relation-screwon}, we turn to the screwon dispersion relation, by which we mean the dependence of energy levels on $\la$ and $k$. We focus attention on a strong coupling regime where the (shifted) energy levels depend on $\la$ and $k$ only through their product. Further restricting $k$ to an intermediate range of wave numbers, ensures that the shifted energies are independent of $p_z$. In this phase, our numerical results show that the shifted energies depend on $\la$ and $k$ through a {\it common power} $\eta(l)$. For $l = 0$, this exponent $\eta(0) = 1$ for the ground state $n = 0$; it rapidly drops to  $\eta(0) = 2/3$ with increasing radial quantum number $n \gtrsim 100$. This $2/3$ power law for radially excited screwons continues to hold as long as $|l| \ll n$ although there are deviations when $|l|$ is comparable to or exceeds $n$. These results confirm and extend our earlier results on the screwon dispersion relation obtained via a WKB approximation in \cite{G-V-3}.  We conclude in Section \ref{s:discussion} with a discussion of possible physical implications of our results and mention some interesting outstanding problems.

\section{Variational approximation to screwon ground state}
\label{s:variational-principle}

\begin{figure}[h]
	\begin{center}
		\centering
		\includegraphics[width=7cm]{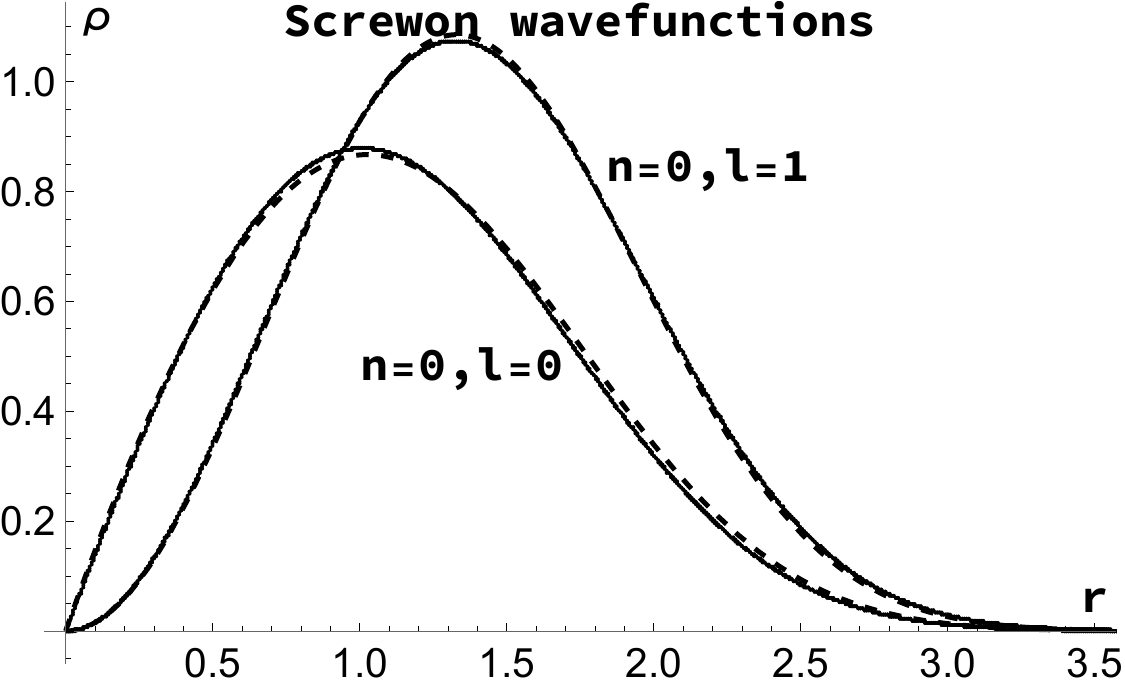}
		\end{center}
	\caption{\footnotesize Comparison of variational (dashed) and finite-difference numerical (solid) radial wavefunctions $\rho(r) = \sqrt{r} \varrho(r)$ for $n =0$ and $l=0,1$ with $\la = k =1$ and $\hbar = m = \mu = 1$. They go to zero as $r$ and $r^2$ as $r \to 0$ and as $e^{-r^3/6}$ as $r \to \infty$.}
	\label{f:GS-Wavefunctions}
	\end{figure}

We may obtain an approximation to the screwon ground state energy using the Rayleigh-Ritz variational principle. From \S 4.3 of \cite{G-V-3}, away from the weak coupling limit, the asymptotic behaviour of the radial wavefunction (\ref{e:product-wavefunction}) is: 
	\beq
	\varrho(r) \sim r^{-1}\exp{\left( -\frac{\sqrt{2 \mu \beta}}{\hbar} \left( \frac{r^3}{3} + \frac{\al r}{2 \beta} \right) \right)} \quad \text{as} \quad r \to \infty \quad \text{and} \quad \varrho(r) \to r^{|l|+ \half} \quad \text{as} \quad r \to 0.
	\label{e:large-r-behav-radial-wfn}
	\eeq
An ansatz that incorporates this asymptotic behavior and has no nodes (as appropriate to the lowest-lying state $n=0$ for any fixed $l$) is 
	\beq
	\varrho_{\rm try}(r) \propto \frac{r^{|l|+\half}}{1 + \zeta r^{|l|+ \frac{3}{2}}} \exp{\left( -\frac{\sqrt{2 \mu \beta}}{\hbar} \left( \frac{r^3}{3} + \frac{\al r}{2 \beta} \right) \right)},
	\label{e:wavefunction-ansatz}
	\eeq
where $\zeta \geq 0$ is a variational parameter (see Fig.~\ref{f:GS-Wavefunctions}). The peak in this wavefunction drops in height and shifts towards $r = 0$ as $\zeta$ increases. Note that this trial wavefunction is not valid in the weak coupling limit since when $\la \to 0$, $\beta \to 0$ and 
	\beq
	\varrho_{\la = 0}(r) \propto r^{|l| + 1/2} \exp\left(-\frac{\sqrt{\mu} k r^2}{2\hbar} \right) L_p^{|l|}\left(-\frac{\sqrt{\mu} k r^2}{\hbar}\right).
	\eeq
Here, $L_p^{|l|}$ are generalized Laguerre polynomials \cite{Pauli} and $p = (n - |l|)/2$. Extremizing the expectation value of the radial Hamiltonian (\ref{e:radial-equation-dimensionful-varrho}) in the state (\ref{e:wavefunction-ansatz}):
	\beq
	E^{\rm var} = \min_{\zeta} \frac{\bra \varrho | H | \varrho \ket}{\bra \varrho | \varrho \ket}
	\quad \text{where} \quad \bra \varrho_1 | \varrho_2 \ket = \int dr \varrho_1^{*}(r) \varrho_2(r),
	\eeq
we find a variational upperbound for the screwon ground state energy. For instance, when $\la = k  = l = \mu = m = p_z = \hbar = 1$, we find the optimal value $\zeta = 0.12$, corresponding to the ground state energy estimate $E^{\rm var}_{n=0,l=1} = 3.416$ which compares favorably with the numerical result $E_{0,1} = 3.415$ from \S \ref{s:finite-difference method}. Similarly, for $l=0$,  $E^{\rm var}_{n=0,l=0} = 1.866$, which is just above the numerical result $E_{0,0} = 1.862$. Moreover, for fixed $k$, the variational ground state energy increases roughly linearly with $\la$ while the optimal value of $\zeta$ decreases. To examine spectral statistics of screwons and their dispersion relation, we need to treat excited states and reliably extract the dependence of energy eigenvalues on parameters. This requires us to go beyond this variational approximation, which we do via numerical diagonalization of $H$ in \S \ref{s:finite-difference method}.
	
\section{First order perturbation theory}
\label{s:First-order-perturbation-theory}

For weak coupling $\la$, we may use perturbation theory to estimate energy eigenvalues. For fixed $p_z$, we write the 2D Hamiltonian (\ref{e:Hamiltonian-quadratic-quartic-Cartesian}) as
	\beqs
	H_{2D} &=& \frac{1}{2 \mu} \left( p_r^2 + \frac{p_\tht^2 - \frac{\hbar^2}{4}}{r^2} \right) + \frac{k^2 r^2}{2} + \frac{p_z^2}{2 \mu} + \frac{k^2 m^2}{2} +\la \left( \frac{k m p_\tht}{2 \mu} - \frac{k p_z r^2}{2 \mu} \right) \cr
	&&+ \la^2 \left(\frac{k^2 m^2 r^2}{8 \mu} + \frac{k^2 r^4}{8 \mu} \right) = H_0 + \la H_1 + \la^2 H_2.
	\eeqs
Note that $H_0$ depends on $p_\tht = L_z$ only through its square. The eigenvalues of the unperturbed Hamiltonian $H_0$ are 
	\beq
	E_{n,l}^{(0)} = (2 n + |l| + 1) \frac{\hbar |k|}{\sqrt{\mu}} + \frac{p_z^2}{2 \mu} + \frac{k^2 m^2}{2}.
	\label{e:eigenvalues-unperturbed}
	\eeq
Here $n =0,1,2, \cdots$ and $l$ is an integer. For $n = 0$, the corresponding normalized eigenvectors of $H_0$ are
	\beq
	\psi_{0,l}^{(0)}(r,\tht) = \left[\left(\frac{\sqrt{\mu} k}{\hbar}\right)^{|l|+1} \frac{1}{\pi |l|!} \right]^{\half} r^{|l|} \exp \left(-\frac{\sqrt{\mu} k}{\hbar} \frac{r^2}{2} \right) \exp(il \tht).
	\eeq
Thus, to first order in perturbation theory, $E_{0,l} = E_{0,l}^{(0)} + \la E_{0,l}^{(1)} + \cdots$ where
	\beq
	E_{0,l}^{(1)} = \bra \psi_{0,l}^{(0)} | H_1| \psi_{0,l}^{(0)} \ket = \frac{k m l \hbar}{2 \mu} - \frac{k p_z}{2 \mu} (|l|+1)\left(\frac{\hbar}{\sqrt{\mu} k} \right).
	\eeq
For instance, if $\hbar = \mu = k = m =  p_z = l=1$, we find the ground state energy
	\beq
	E_{0,0} = E_{0,0}^{(0)} + \la E_{0,0}^{(1)} + {\cal O}(\la^2) = 2 - \frac{\la}{2} + {\cal O}(\la^2).
	\eeq
For $\la = 0.01, 0.05, 0.08$, the ground state energies $E_{0,0}$ with first order correction are $1.995, 1.975$ and $1.96$ while those obtained numerically via finite differences are $1.99,1.952$ and $1.925$, showing the loss in accuracy of first order perturbation theory with increasing $\la$. Nevertheless, for small $\la$, perturbation theory predicts a drop in the energy with increasing $\la$. This is visible in the numerical results shown in Fig.~\ref{f:E-n-l-vs-Lambda}.

\section{Screwon spectrum via finite-differences}
\label{s:finite-difference method}

Here, we address the question of determining the spectrum of screwons by numerical diagonalization of the Hamiltonian (\ref{e:Hamiltonian-quadratic-quartic-Cartesian}) which (aside from free particle motion along $z$) describes a cylindrically symmetric anharmonic oscillator in the $x$-$y$ plane. A direct approach would be to use finite differences to discretize the 2D Hamiltonian, truncate it and diagonalize the resulting matrix. An alternative is to numerically determine the spectrum of the radial Hamiltonian (\ref{e:radial-equation-dimensionful-varrho}) in each angular momentum sector ($l = 0,1,2, \ldots$) and then combine these sectors to obtain the full screwon spectrum. We implemented both approaches and found the second to be more appropriate for the following reasons. (a) For comparable precision, diagonalizing the 2D Hamiltonian was computationally  slower. (b) A 2D square lattice does not possess the circular symmetry of $H$ making it harder to accurately  capture the behavior of the wave function as $r \to 0$ and for large $r$. (c) We wish to classify energy levels using the quantum numbers $n$ (number of nodes of $\varrho(r)$) and $l$ to facilitate the search for level crossings.  (d) We need the spectrum of the radial Hamiltonian for fixed $l$ in order to look for a power law dispersion relation $E_{nl} \sim (\la k)^\eta$.

To find the spectrum of (\ref{e:radial-equation-dimensionful-varrho}), we discretize the radial coordinate $r = i \delta$ for $i = 1,2, \ldots, N_r$, with a small spacing $\delta$. This converts (\ref{e:radial-equation-dimensionful-varrho}) into an eigenvalue problem for an $N_r \times N_r$ matrix. However, direct implementation using the wavefunction defined in (\ref{e:product-wavefunction}) $\psi \propto \varrho/ \sqrt{r}$ leads to eigenvalues that converge very slowly (with increasing $N_r$) to the known spectrum (\ref{e:eigenvalues-unperturbed}) for $\la = 0$ when $l = 0$. We traced the numerical difficulty to the behavior of $\varrho(r)$ as $r \to 0$. From \cite{G-V-3} it is known that
	\beq
	\varrho(r) \to r^{|l| + \half} \quad \text{as} \quad r \to 0.
	\eeq
We then found that the numerical scheme converges much faster if we use (in place of (\ref{e:product-wavefunction})), the radial wavefunction 
	\beq
	\psi_a(r, \tht, z) = \frac{1}{r^a} \rho(r) \exp{(i l \tht)} \exp{\left(\frac{i p_z z}{\hbar}\right)} \quad \text{with} \quad a = 1.
	\label{e:wavefunction-ansatz-generic}
	\eeq
This ensures a softer behaviour of the radial wavefunction near the origin: $\rho(r) \to r^{a +|l|}$ as $r \to 0$. The radial equation corresponding to (\ref{e:wavefunction-ansatz-generic}) is
	\beq
	-\frac{\hbar^2}{2 \mu} \left( \rho''(r) + \frac{1-2a}{r} \rho' + \frac{a^2 -l^2}{r^2} \rho \right) + \left( \al r^2 + \beta r^4 + \frac{\hbar l \la k m}{2 \mu} + \frac{p_z^2}{2 \mu} + \frac{k^2 m^2}{2} \right) \rho = E \rho.
	\label{e:radial-equation-generic}
	\eeq
As mentioned, we discretize $r$ with equal spacing $r_i = i \delta$ for $i = 1,2, \ldots, N_r$. To ensure that the discretized Hamiltonian is real symmetric, we use the centered-difference formulae:
	\beq
	\rho_i'' = \frac{\rho_{i+1} - 2 \rho_i + \rho_{i-1}}{\delta^2}  \quad \text{and} \quad \rho_i' = \frac{\rho_{i+1} - \rho_{i-1}}{2 \delta}.
	\eeq
The radial equation becomes the system of difference equations:
\small
	\beqs
	&&-\frac{\hbar^2}{2 \mu} \left(\frac{\rho_{i+1} - 2 \rho_i + \rho_{i-1}}{\delta^2} +   \frac{(1-2a)(\rho_{i+1} - \rho_{i-1})}{2 i \delta^2} \right)  \cr
	&&+  \left(  \frac{\hbar^2}{2 \mu i^2 \delta^2} \left( l^2 - a^2 \right) + \al i^2 \delta^2 + \beta i^4 \delta^4 + \frac{\hbar l \la k m}{2 \mu} + \frac{p_z^2}{2 \mu} + \frac{k^2 m^2}{2} \right) \rho_i = E \rho_i. 
	\eeqs
\normalsize
We read off the discretized radial Hamiltonian, which is an $N_r \times N_r$ tri-diagonal matrix with nontrivial entries
\small
	\beqs
	H^{r}_{ii} &=& \frac{\hbar^2}{\mu \delta^2} + \alpha i^2 \delta^2 + \beta i^4 \delta^4 + \frac{\hbar^2}{2 \mu i^2 \delta^2} \left( l^2 - a^2 \right) + \frac{\hbar l \la k m}{2 \mu} + \frac{p_z^2}{2 \mu} + \frac{k^2 m^2}{2} \quad \text{and} \cr
	H^{r}_{i, i \pm 1} &=& -\frac{\hbar^2}{2 \mu \del^2} \pm \frac{(1-2a)}{2 i \del^2}. \quad
	\eeqs
\normalsize
The `boundary' matrix elements are $H_{1,0} = H_{N_r, N_r+1} = 0$. It may also be written as
\small
	\beqs
	H^{\rm r}_{ij} &=& \left[- \frac{\hbar^2}{2 \mu} \left( -\frac{2}{\delta^2} + \frac{a^2 - l^2}{i^2 \delta^2} \right) + \al i^2 \delta^2 + \beta i^4 \delta^4 + \frac{\hbar l \la km}{2\mu}  + \frac{p_z^2}{2 \mu} + \frac{k^2 m^2}{2}  \right] \delta_{ij} \cr
	&& - \frac{\hbar^2}{2 \mu} \left( \frac{1}{\delta^2} + \frac{1-2a}{2 i \delta^2} \right) \delta_{i, j-1} - \frac{\hbar^2}{2 \mu} \left( \frac{1}{\delta^2} - \frac{1-2a}{2 i \delta^2} \right) \delta_{i, j+1} 
	\quad \text{for} \quad 1 \leq i,j \leq N_r.
	\label{e:rad-ham-fin-dif-mat-ele}
	\eeqs
\normalsize
Our aim is two-fold: (i) to get the first $N_l$ energy levels $E_{n,l}$ for each of several values of $l$ with $|l| \leq l_{\rm max}$ and (ii) to combine these spectra to get the first $N$ energy levels accounting for all values of $l$. The latter is facilitated by the fact that the lowest energy for a given $l$ increases with $|l|$.

Both (i) and (ii) are to be done for various values of $\la$, $k$  and $p_z$ for fixed $\mu =  m = \hbar = 1$, by a choice of units. To find the lowest $N_l$ levels of the radial Hamiltonian $H^r_l$, we need to pick a spacing $\del$ for the radial grid $r_i = i \del$ for $i = 1, \cdots, N_r$ to construct the $N_r \times N_r$ matrix approximation to $H^r$ (\ref{e:rad-ham-fin-dif-mat-ele}). For fixed $N_r$, $\del$ must be chosen so that the first $N_l$ radial eigenfunctions $\psi_{n,l}(r)$ are accurately captured. In particular, $r_{\rm max} = N_r \del$ must be large enough to accommodate all the oscillations in the most highly excited ($N_l^{\rm th}$) state desired, which must decay as $e^{- \text{const} \times r^3}$ as $r \to \infty$ (\ref{e:large-r-behav-radial-wfn}). On the other hand, $\del$ must be small enough to resolve each of the $N_l$ oscillations of this wavefunction. From (\ref{e:large-r-behav-radial-wfn}), the optimal value of $\del$ is a decreasing function of both $k$ and $\la$. Moreover, due to truncation errors, we may trust only a small fraction $N_l/N_r \ll 1$ of the $N_r$ energy levels computed by diagonalizing (\ref{e:rad-ham-fin-dif-mat-ele}). In practice, for $\la = k = p_z = \mu = m = 1$ we find that we are able to reliably obtain the lowest $N_l = 100$ levels by choosing $N_r = 5000$ and $r_{\rm max} = 25$. In this case, we find that the lowest $N_l$ energy eigenvalues change by less that $0.1 \%$ if the size of the matrix $N_r$ is increased from 4000 to 5000. Moreover, for $N_l = 100$ we find that the highest energy for $l = 0$ is $E_{\rm max} = E_{N_l,0} = 803.507$ while the ground state energy in all angular momentum sectors with $|l| > 200$ lies above this value. Thus, by merging the spectra for $E \leq E_{\rm max}$ from all sectors with $|l| < l_{\rm max} = 200$, we obtain the first $N = 21061$ screwon energy levels. We have verified for a few values of $l$ (and $\la = 1$) that the ground state energies obtained numerically lie marginally below the variational upper bounds obtained in \S \ref{s:variational-principle}.

\section{Energy level statistics}
\label{s:energy-level-statistics}
	
Energy levels $E_{n,l}$ of the radial Hamiltonian $(\ref{e:radial-equation-dimensionful-varrho})$ may be labelled by a principal quantum number $n = 0,1,2, \ldots$ (the number of nodes of the radial wavefunction $\varrho(r)$ in $(0, \infty)$) and by the angular momentum quantum number $l = 0, \pm 1, \pm 2, \ldots$. For $\la = 0$, the spectrum is given by \cite{G-V-3} 
	\beq
	E_{n, l} = (2n + |l| + 1) \frac{\hbar |k|}{\sqrt{\mu}} + \frac{p_z^2}{2 \mu} + \frac{k^2 m^2}{2}.
	\eeq
Working in units where $\hbar = m = \mu =1$ and taking for definiteness $k = p_z = 1$, gives $E_{n,l} = 2n + |l| + 2$. The first few levels are as tabulated below:
	\begin{center}
		\begin{tabular}{|c|c|c|c|c|c|c|c|c|c|c|} 
	 	\hline
 		$E$ & 2 & 3 & 3 & 4 & 4 & 4  & 5 & 5 & 5 & 5 \\ 
		\hline
		$n$ & 0 & 0 & 0 &  0 & 0 & 1 & 0 & 0 & 1 & 1 \\ 
		\hline
 		$l$ & 0 & 1 & -1 & 2 & -2 & 0   & 3 & -3 & 1 & -1 \\ 
 		\hline
		\end{tabular}
	\end{center}
For these parameter values, the level with energy $E$ is $(E-1)$-fold degenerate. More generally, the level with energy $E_{n,l}$ is $(2n + |l| +1)$-fold degenerate. 

\begin{figure}[h]
	\begin{center}
		\begin{subfigure}[t]{7cm}
		\centering
		\includegraphics[width=7cm]{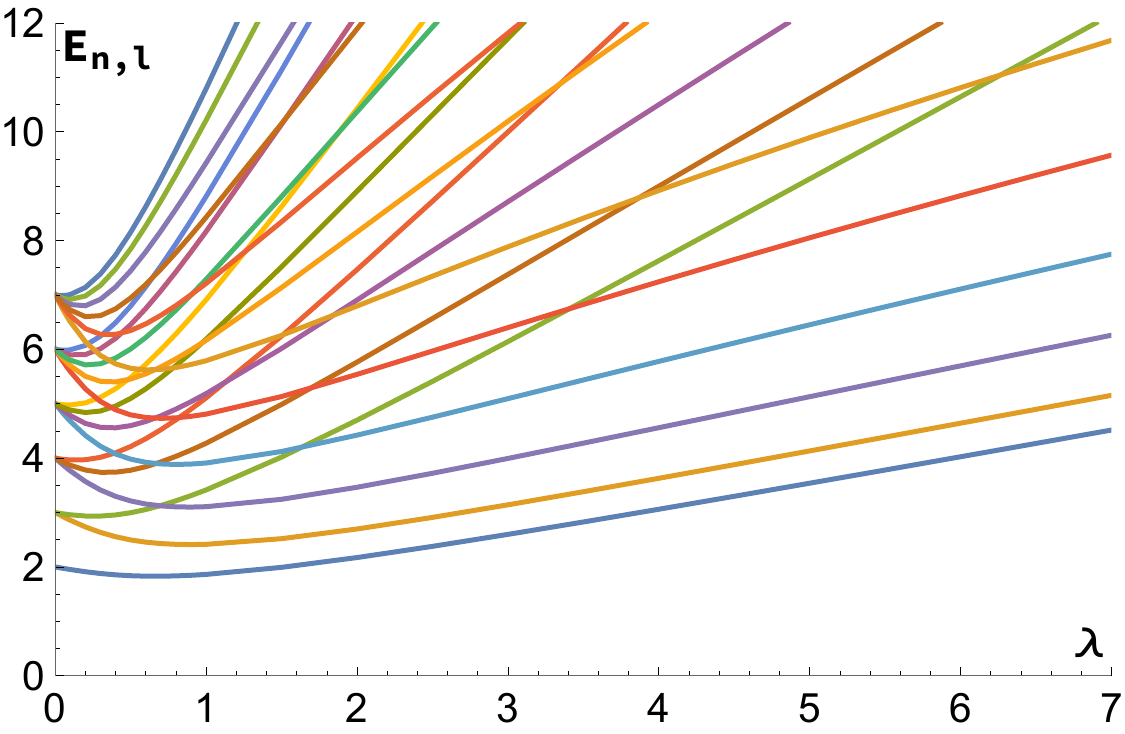}
		\caption{}
		\label{f:E-vs-lambda}
		\end{subfigure}
		\qquad
		\begin{subfigure}[t]{7cm}
		\centering
		\includegraphics[width=7cm]{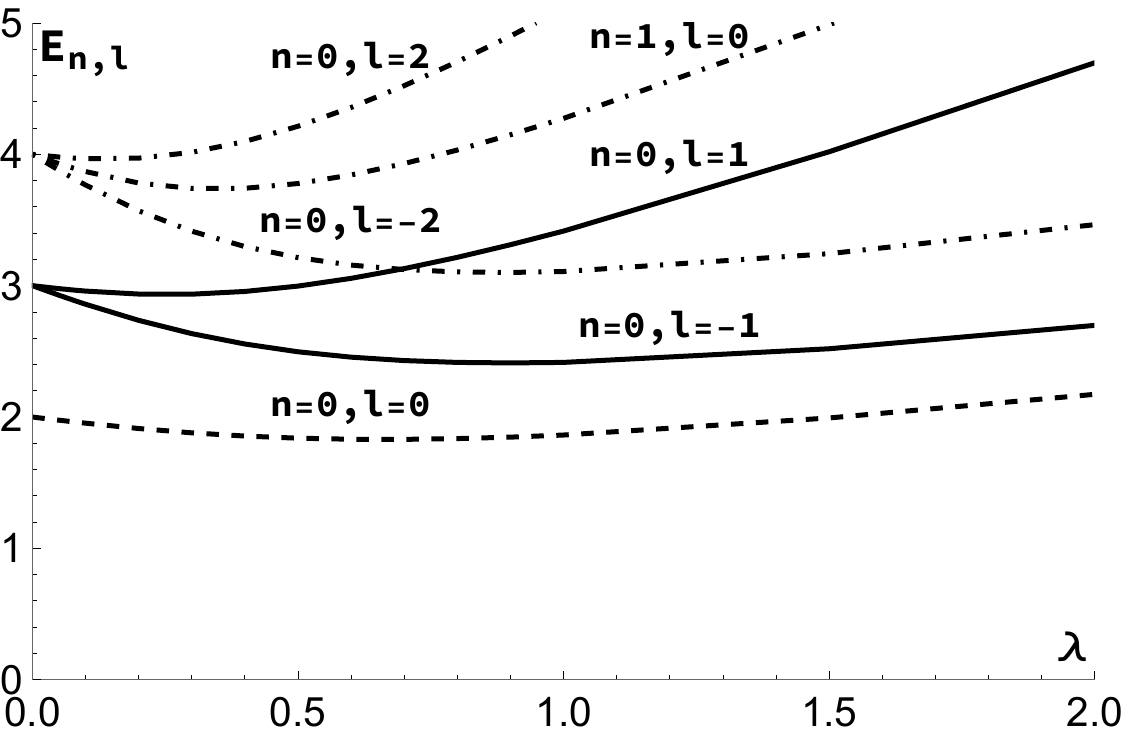}
		\caption{}
		\label{f:E-vs-lambda-enlarged}
		\end{subfigure}
		\end{center}
	\caption{\footnotesize Lowest lying screwon energy levels $E_{nl}$ vs coupling $\la$ for $k = p_z = \mu = m = 1$. Degeneracies present when $\la = 0$ break for $\la > 0$ though accidental degeneracies occur at level crossings. The initial drop in energy with increasing $\la$ is explained by 1st order perturbation theory in \S \ref{s:First-order-perturbation-theory}.}
	\label{f:E-n-l-vs-Lambda}
	\end{figure}

For $\la > 0$, we find that these degeneracies are generally lost. Heuristically, the rotational energy $\hbar l \la k m/2 \mu$ in (\ref{e:radial-equation-dimensionful-varrho}) suggests that within a $\la =0$ degenerate multiplet, levels with higher $l$ acquire larger energies. This is illustrated in Fig.~\ref{f:E-n-l-vs-Lambda}, which also shows that there are no avoided level crossings. The resulting accidental degeneracies and the lack of level repulsion are typical features of integrable systems.

\subsection{Semiclassical cumulative energy level distribution}
\label{s:semiclass-density-states}

\begin{figure}[h]
	\begin{center}
		\begin{subfigure}[t]{6.3cm}
		\centering
		\includegraphics[width=6.3cm]{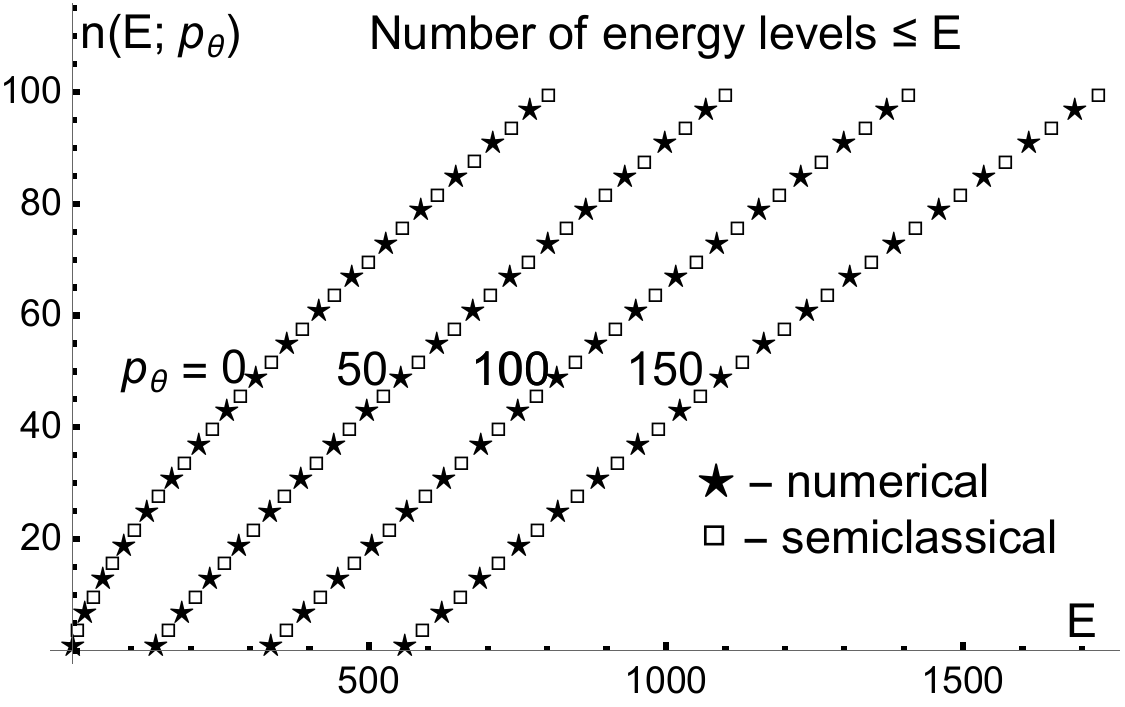}
		\caption{}
		\label{f:semiclass-cdf-compare-fix-ptht}
		\end{subfigure}
		\qquad
		\begin{subfigure}[t]{6.3cm}
		\centering
		\includegraphics[width=6.3cm]{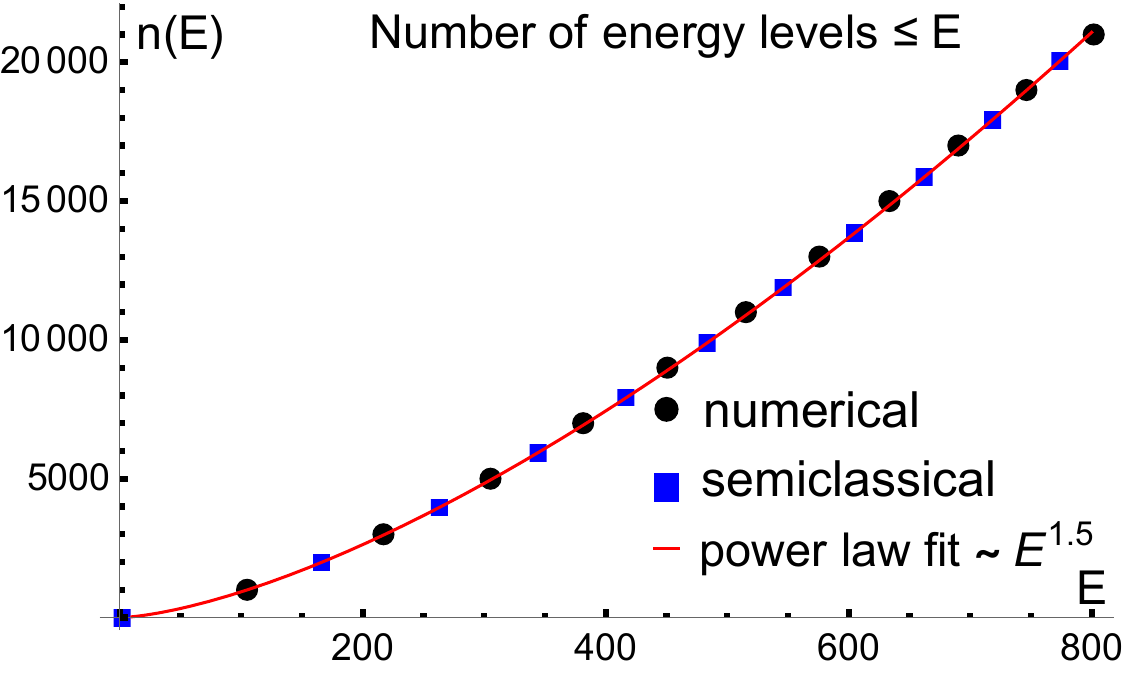}
		\caption{}
		\label{f:semiclass-cdf-compare-all-ptht}
		\end{subfigure}
		\end{center}
		\caption{\small (a) Semiclassical cumulative distribution $n(E, \hbar; p_\tht)$  (squares) nearly matches that from numerical diagonalization (stars) for $p_\tht = 0, 50, 100, 150$. (b) Comparison of semiclassical (squares) and numerical (disks) $n(E,\hbar)$ after combining all allowed values of $p_\tht$. In all cases we take $p_z = 1$ and $k = \mu = m = \la = \hbar = 1$.}
		\label{f:semiclass-cdf-compare}
	\end{figure}

Here, we obtain a semiclassical estimate for the cumulative energy level distribution $n(E,\hbar)$ and compare it with that from numerical diagonalization. As in \S \ref{s:First-order-perturbation-theory}, we fix $p_z$ and restrict attention to the $x$-$y$ plane so that the energy spectrum $E_i$ is discrete. Recall that $n(E,\hbar)$ is the number of levels with energy less than or equal to $E$:
	\beq
	n(E, \hbar) = \sum_{i = 1}^{N} \tht(E- E_i).
	\label{e:cdf-n-of-E-hbar}
	\eeq 
Here, $\theta(x)$ is the unit step function defined as: $\theta(x) = 1$ for $x \geq 0$ and $\theta(x) = 0$ for $x < 0$. If we suppose that each state occupies a volume $h^2$ in the 4D phase space, then
	\beq
	n(E,\hbar) \sim \ov{(2\pi \hbar)^2} \int_{H \leq E} d x \, dy \, dp_x \, dp_y
	= \ov{(2\pi \hbar)^2} \int_{H \leq E} d r \, d\tht \, dp_r \, dp_\tht.
	\label{e:semiclass-cdf}
	\eeq
The transformation to polar coordinates $r = \sqrt{x^2 + y^2}$, $\tht = \arccos(x/r)$, $p_r = (x p_x + y p_y)/r$ and $p_\tht = x p_y - y p_x$ is canonical, so the volume element is preserved. Equation (\ref{e:semiclass-cdf}) is expected to hold in the semiclassical regime of small $\hbar$ or large $E$. [For the classical $\la = 0$ model, the combinations of system parameters (i.e., not including $p_z$ and $p_\tht$) with dimensions of action and energy are $k m^2 \sqrt{\mu}$ and $k^2 m^2$, so semiclassical would mean $\hbar \ll k m^2 \sqrt{\mu}$ or $E \gg k^2 m^2$. For $\la \ne 0$, there are other such combinations due to the presence of the dimensionless parameter $\sqrt{\mu}/m \la$]. The classical 2D Hamiltonian (\ref{e:Hamiltonian-quadratic-quartic-Cartesian}) (for fixed $p_z$)
	\beq
	H = \frac{p_r^2}{2\mu} + \frac{p_\tht^2}{2 \mu r^2} + \al r^2 + \beta r^4 + \frac{\la k m p_\tht}{2 \mu}  + \frac{p_z^2}{2 \mu} + \frac{k^2 m^2}{2}
	\label{e:2d-hamiltonian-polar}
	\eeq
admits $\tht$ as a cyclic coordinate so that $p_\tht$ is conserved. The integral over $\tht$ gives $2\pi$. We may choose to do the $p_\tht$ integral last:
	\beq
	n(E,\hbar) \sim \ov{(2\pi \hbar)^2} \int_{p_\tht^{\rm min}(E)}^{p_\tht^{\rm max}(E)} d p_\tht \int_0^{2\pi} d\tht \int_{H_{p_\tht}(r,p_r) \leq E} dr \, d p_r 
	= \int_{p_\tht^{\rm min}(E)}^{p_\tht^{\rm max}(E)} n(E, \hbar; p_\tht) \, d p_\tht.
	\label{e:num-lvl-less-E-sum-over-fix-ptht}
	\eeq
Here $H_{p_\tht}(r,p_r)$ is the Hamiltonian (\ref{e:2d-hamiltonian-polar}) for fixed $p_\tht$ and
	\beq
	n(E, \hbar; p_\tht) \sim \ov{2\pi \hbar^2} \int_{H_{p_\tht}(r,p_r) \leq E} d r \, dp_r
	\label{e:num-lvl-less-E-fix-ptht}
	\eeq
may be interpreted as the semiclassical number of states of given $p_\tht$ with $H_{p_\tht} \leq E$. The limits of integration $p_\tht^{\rm min,max}(E)$ are obtained by extremizing with respect to $r$ and $p_r$ in
	\beq
	p_\tht^{\pm}(r,p_r;E) = \frac{r}{2} \left[- k m \la r \pm \sqrt{ - 4 p_r^2 - (\la k r^2 - 2 p_z)^2 + 8 E \mu - 4 k^2 \mu (m^2 + r^2) } \right].
	\eeq
The expressions for $p_\tht^\pm$ are got by solving the quadratic equation $H = E$ for $p_\tht$. The extrema occur at $p_r = 0$ and values of $r$ determined via the roots of a quartic equation in $r^2$:
	\beqs
	&& 9 k^4 \la^4 r^8 
	- 4 k^3 \la^2 (12 \la p_z - k \la^2 m^2 - 12 k \mu) r^6 \cr
	&& - 8 k^2 [6 E \la^2 \mu -11 \la^2 p_z^2 + 2 k p_z \la (m^2 \la^2 + 8 \mu) - k^2 \mu (5 m^2 \la^2 + 8 \mu) ] r^4 \cr
	&& +16 k (k m^2 \la^2 + 4 k \mu -4 p_z \la)(p_z^2 - 2 E \mu + k^2 m^2 \mu)r^2 
	+ 16 (p_z^2 - 2 E \mu + k^2 m^2 \mu)^2 = 0. \quad
	\eeqs
We evaluate the integral in (\ref{e:num-lvl-less-E-fix-ptht}) numerically for several energies $E$ for fixed values of $p_\tht, p_z$ and other parameters. The resulting distributions $n(E, \hbar; p_\tht)$ are compared in Fig.~\ref{f:semiclass-cdf-compare-fix-ptht} with those obtained from numerical diagonalization. We find that the semiclassical estimate for $n(E, \hbar = 1; p_\tht)$ agrees quite well with the numerical spectrum for each allowed value of $p_\tht$ and $E \gg k^2 m^2$. To obtain $n(E)$, we numerically integrate over $p_\tht$ in (\ref{e:num-lvl-less-E-sum-over-fix-ptht}). The resulting cumulative distribution is compared with that from numerical diagonalization in Fig.~\ref{f:semiclass-cdf-compare-all-ptht}.

\paragraph*{\bf Power law behavior of $n(E; p_\tht)$ and $n(E)$.} For large $E$, we find that the cumulative level distribution satisfies a power law $n(E; p_\tht) \propto (E - E_0)^{\zeta}$ where $E_0(p_\tht)$ is the ground state energy. Since the term $\la k m p_\tht/2 \mu$ linear in $p_\tht$ is a constant addition to the Hamiltonian (\ref{e:2d-hamiltonian-polar}), it does not contribute to energy differences. Thus, the power $\zeta(p_\tht)$ depends only on $|p_\tht|$. We find that it increases monotonically with $|p_\tht|$. Moreover, after combining all angular momentum sectors, $n(E)$ also satisfies a power law for large $E$. For instance, when $\hbar = k = m = \mu = \la = p_z = 1$, $\zeta(p_\tht) \approx 0.75, 0.78, 0.81, 0.84, 0.88$ for $p_\tht = 0, \pm 20, \pm 50, \pm 100, \pm 150$ while $n(E) \sim E^{1.5}$ for large $E$.

\subsection{Level spacing distributions}

The quantum energy levels of a classically integrable system are expected to display certain universal statistical features such as Poissonian level spacing distributions \cite{B-T}. To remove the system-specific information and extract universal properties\footnote{To study spectral statistics, we note that there is no need to `purify' the screwon spectrum obtained in \S \ref{s:finite-difference method}. Reflection through the origin is part of the circular symmetry of (\ref{e:Hamiltonian-quadratic-quartic-Cartesian}).} of the level spacings, we `unfold' the energy spectrum so that the mean level spacing is one \cite{Haake, G-M-W}. In order to do this, we define unfolded levels $\xi_i = \xi(E_i)$ for $i = 1, 2, 3, \ldots, N$, where $N$ is the number of levels considered ($N \approx 20000$ below). The function $\xi$ is chosen to be the polynomial (typically of degree 5 or 6) that best fits the cumulative energy distribution function introduced in (\ref{e:cdf-n-of-E-hbar}): 
	\beq
	n(E) = \sum_{i = 1}^N \theta(E-E_i) = \xi(E) + \eta(E).
	\label{e:stair-case-function}
	\eeq
Thus, $\xi(E)$ is  a smoothed version of the `staircase function' $n(E)$, which satisfies $n(E_j) =j$ for $j =1, 2, \ldots, N$ assuming there are no degenerate levels. [More generally, if the first $n_1$ energies are degenerate ($E_1 = E_2 = \cdots = E_{n_1}$) and next $n_2$ energies are degenerate ($E_{n_1 +1} = E_{n_1 + 2} = \cdots = E_{n_1+ n_2}$) etc., then $n(E_1) = \cdots = n(E_{n_1}) = n_1$ and 
 $n(E_{n_1+1}) = \cdots = n(E_{n_1+n_2}) = n_1+ n_2$ and so forth.] Here $\eta(E)$ contains fluctuations. By construction, the unfolded levels $\xi_1, \xi_2, \ldots$ have a mean spacing of approximately one. 

\begin{figure}[h]
	\begin{center}
		\begin{subfigure}[t]{5cm}
		\centering
		\includegraphics[width=5cm]{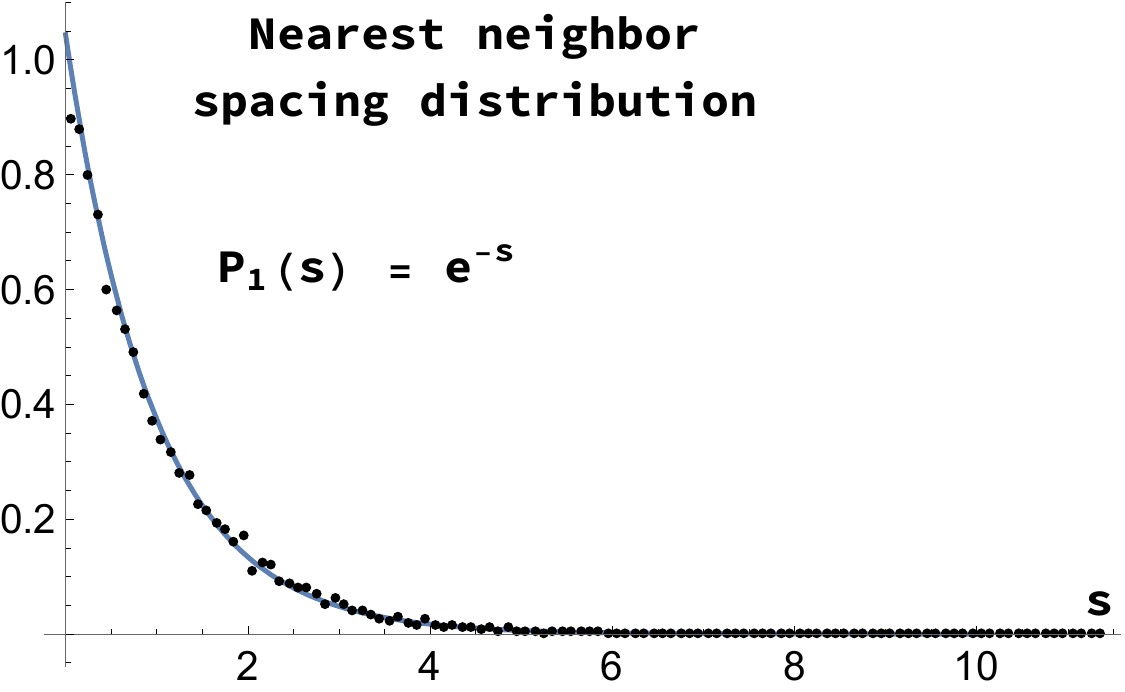}
		\caption{}
		\label{f:nearest-neigh-spacing}
		\end{subfigure}
		\qquad
		\begin{subfigure}[t]{5cm}
		\centering
		\includegraphics[width=5cm]{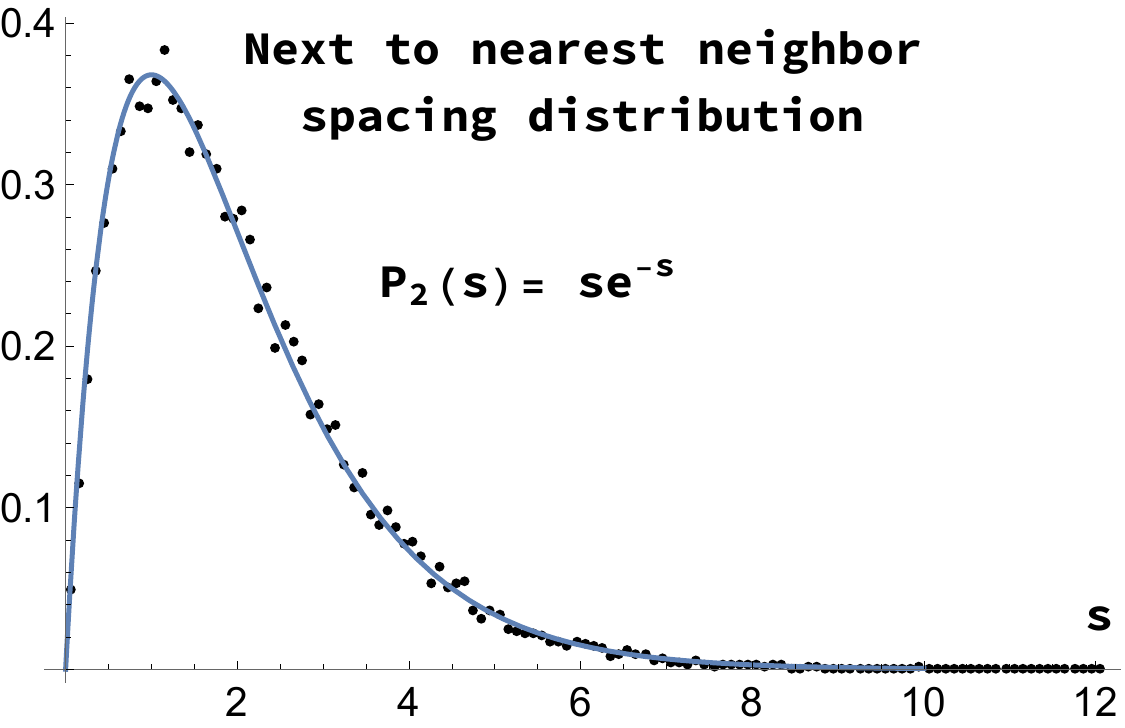}
		\caption{}
		\label{f:nnn-spacing}
		\end{subfigure}
		\qquad
		\begin{subfigure}[t]{5cm}
		\centering
		\includegraphics[width=5cm]{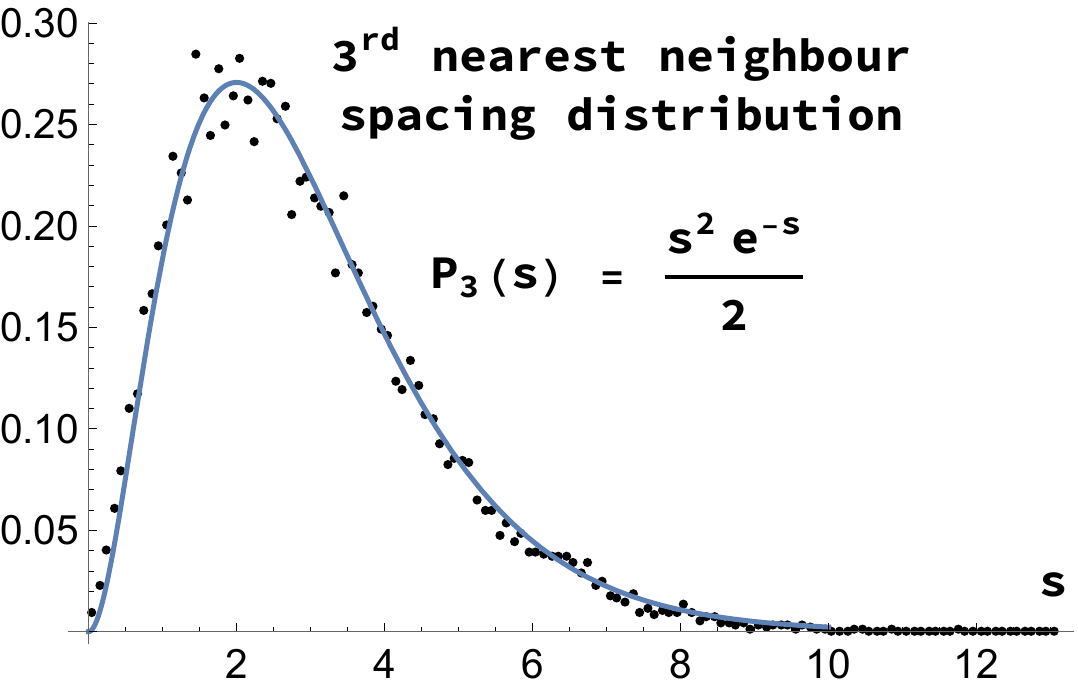}
		\caption{}
		\label{f:spac-3}
		\end{subfigure}
	\end{center}
		\caption{\footnotesize First, second and third order normalized spacing distributions of lowest $N \approx 20000$ unfolded energy levels for $\la = p_z = \mu = k = m = 1$. The solid line is from (\ref{e:k-spacing-distribution}), while the dots represent binheights of the normalized spacing histograms with binwidths 0.2. The spacing distributions are seen to obey Poisson statistics as expected of an integrable system. In fact, we find that Poisson statistics applies for all nonzero coupling $\la$.}
	\label{f:spacing-distributions}
	\end{figure}

For $i = 1, 2, 3, \ldots$, the $i^{\rm th}$-nearest neighbor (n.n.) spacing distribution is defined as the normalized distribution of values of $s = (\xi_{j+i} - \xi_j)$ for $j =1, 2, \ldots N-i$. In Fig.~\ref{f:spacing-distributions}, we plot histograms of the spacing probability distributions for $i = 1,2, 3$. We find that the n.n. spacing distribution follows the exponential law $P_1(s) = \exp(-s)$, while the higher order spacing distributions are well approximated by the Poisson distribution:
	\beq
	P_i(s)  = \frac{s^{i-1} \exp(-s)}{(i-1)!} \quad \text{for} \;\; i = 1,2,3, \ldots.
	\label{e:k-spacing-distribution}
	\eeq
This Poisson statistics is expected if the n.n. spacing is exponentially distributed and successive spacings can be treated as independent random variables.

{\fl \bf Remark:} We find that when we use the unfolded spectrum, the n.n. spacing histograms for all $\la > 0$ have roughly the same range ($0 \leq s \leq 4$). Moreover, the same binwidth (anything from 0.1 to 1) gives the universal exponential distribution for all $\la > 0$. If we do not unfold the spectra, then the mean spacing varies with $\la$ so the spacing distributions have $\la$ dependent ranges. However, it is still possible to extract the universal exponential spacing distribution provided the binwidth is chosen appropriately: a different range of binwidths works for different values of $\la$.

\begin{figure}[h]
	\begin{center}
		\begin{subfigure}[t]{6cm}
		\centering
		\includegraphics[width=6cm]{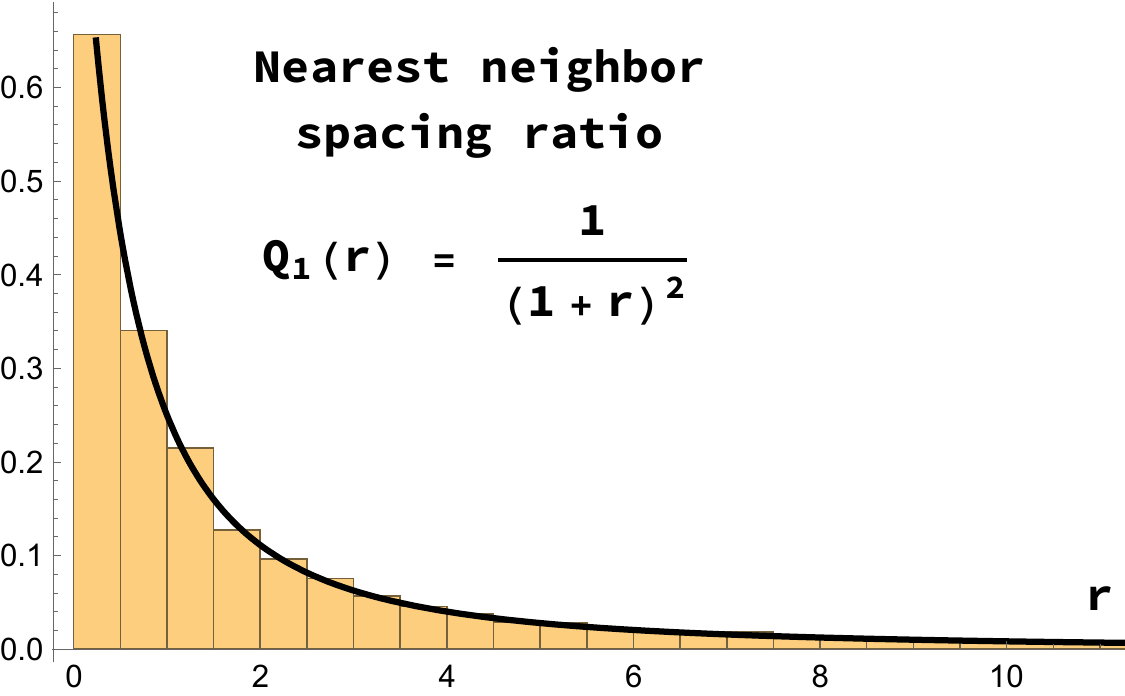}
		\caption{}
		\label{f:spac-ratio-1}
		\end{subfigure}
		\qquad
		\begin{subfigure}[t]{6cm}
		\centering
		\includegraphics[width=6cm]{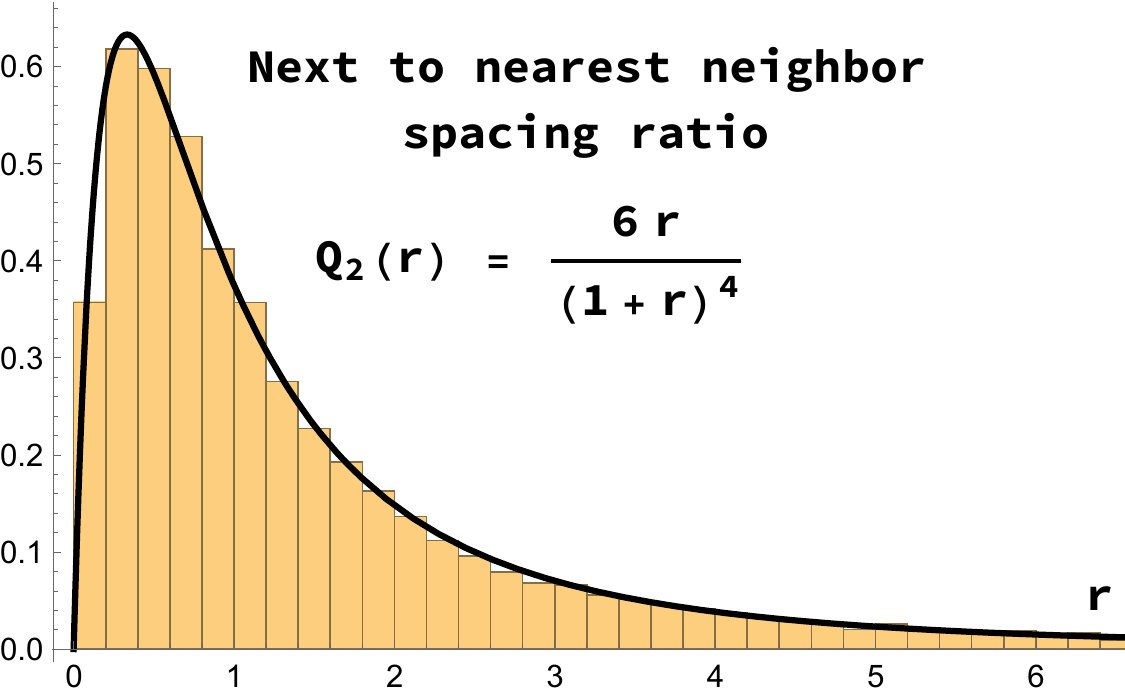}
		\caption{}
		\label{f:spac-ratio-2}
		\end{subfigure}
		\qquad
		\begin{subfigure}[t]{6cm}
		\centering
		\includegraphics[width=6cm]{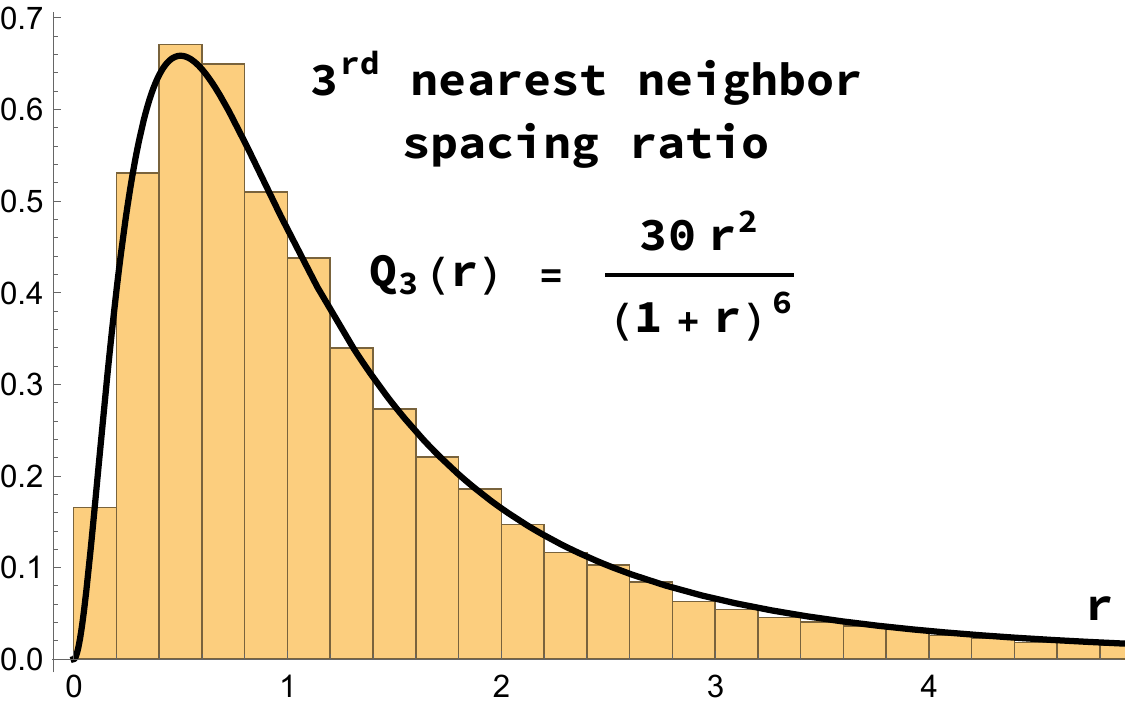}
		\caption{}
		\label{f:spac-ratio-3}
		\end{subfigure}
		\qquad
		\begin{subfigure}[t]{6cm}
		\centering
		\includegraphics[width=6cm]{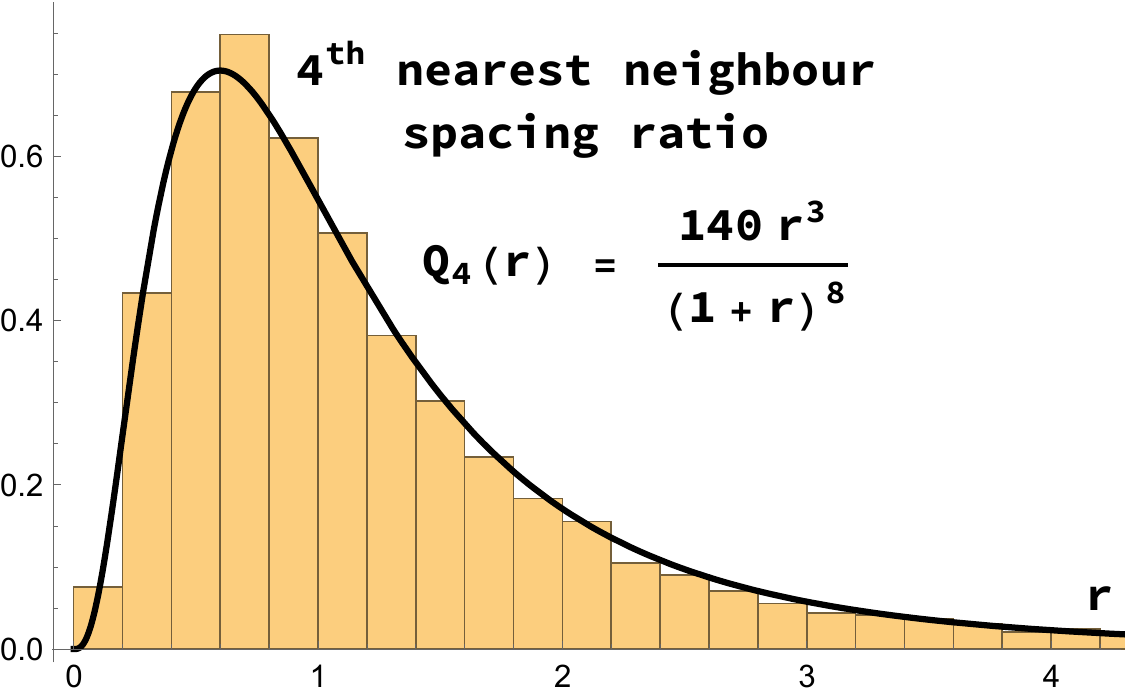}
		\caption{}
		\label{f:spac-ratio-4}
		\end{subfigure}
	\end{center}
		\caption{\footnotesize Histograms of $i^{\rm th}$ nonoverlapping spacing ratio distributions for $i = 1,2,3,4$ for lowest 20000 screwon energy levels with $\la = p_z = k =  \mu = m = 1$. They are seen to match the distributions $Q_i(r)$ (\ref{e:spacing-ratio-distributions}) denoted by solid curves. Unfolding the spectrum does not significantly affect these histograms.}
	\label{f:spacing-ratio}
	\end{figure}

Recently, ratios of level spacings have emerged as a useful statistical tool to study quantum spectra \cite{T-K-S, T-B-S}. The $i^{\rm th}$ nonoverlapping spacing ratio distribution is defined as the normalized distribution of values of
	\beq
	r_j^{(i)} = \frac{E_{j+i} - E_{j}}{E_{j-1} - E_{j-1-i}} \quad \text{for} \quad  j = i+2, \ldots, N-i.
	\eeq
If one assumes that the nonoverlapping spacings are independently distributed, then the probability that the spacing in the numerator is $s'$ and that in the denominator is $s$, is simply the product of individual spacing probabilities. Thus, the probability that the ratio of spacings is $s'/s = r$ is 
	\beq
	Q_{i}(r)  = \iint P_i(s') P_i(s) \delta\left(\frac{s'}{s} - r\right) ds \; ds' = \int P_i(rs) P_i(s) |s| \: ds  = \frac{(2i-1)!}{[(i-1)!]^2} \frac{r^{i-1}}{(1+r)^{2i}}.
	\label{e:spacing-ratio-distributions}
	\eeq 	
This is compared with our numerical results in  Fig.~\ref{f:spacing-ratio}.

\subsection{Number variance}
\label{s:number-variance}

\begin{figure}
	\begin{center}
		\begin{subfigure}[t]{7cm}
		\centering
		\includegraphics[width=7cm]{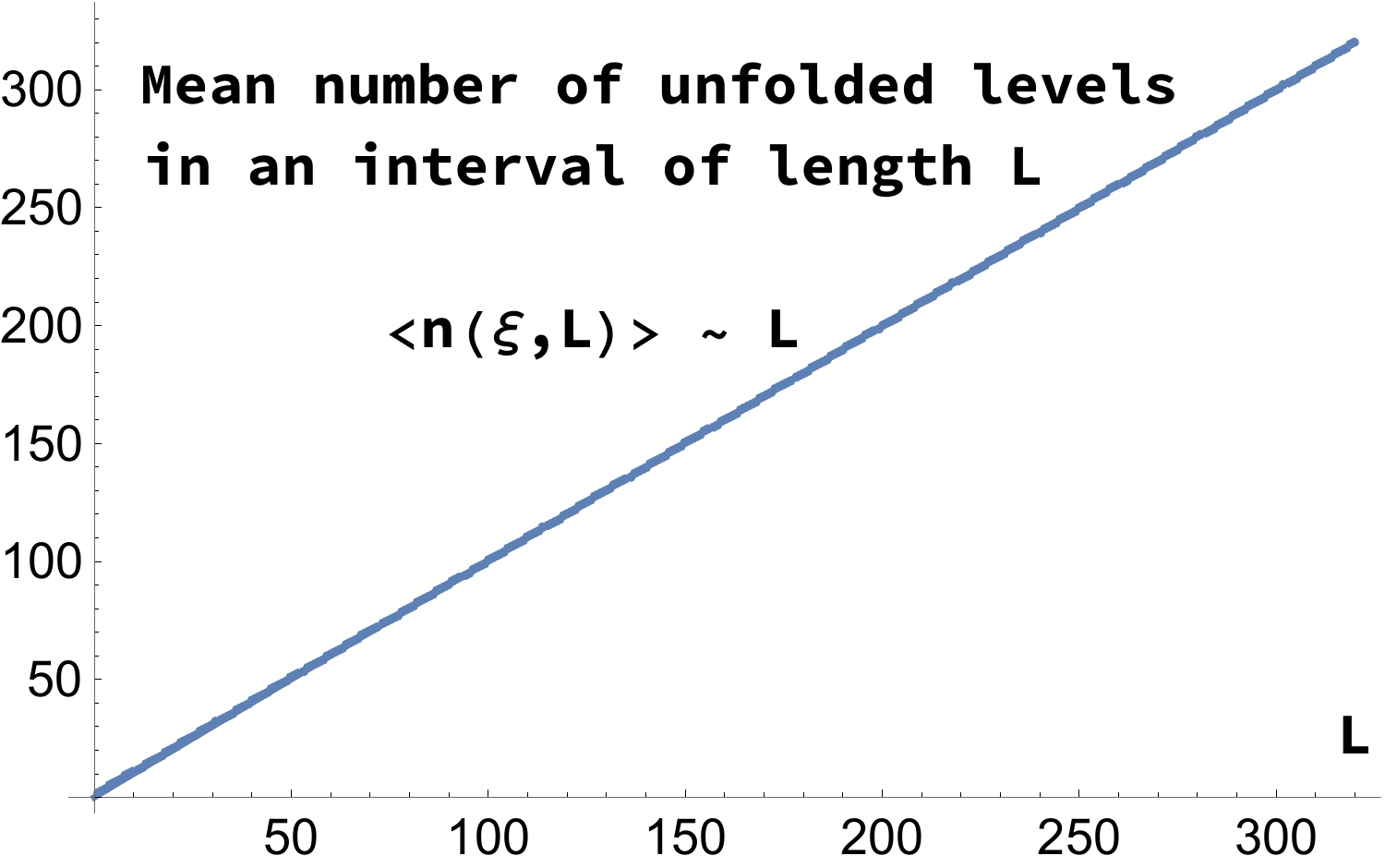}
		\caption{}
		\label{f:num-avg}
		\end{subfigure}
		\qquad
		\begin{subfigure}[t]{7cm}
		\centering
		\includegraphics[width=7cm]{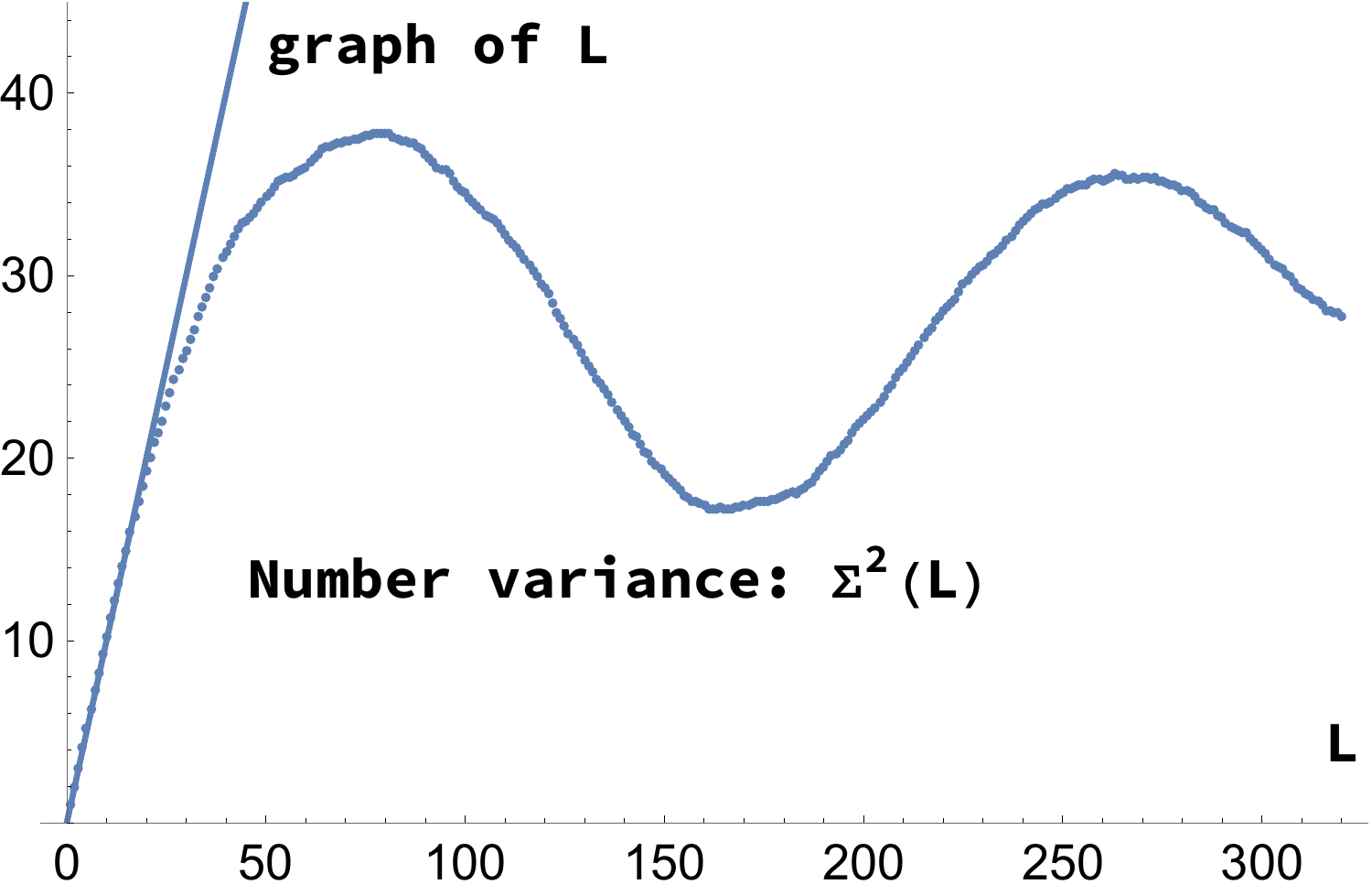}
		\caption{}
		\label{f:num-variance}
		\end{subfigure}
	\end{center}
	\caption{\footnotesize (a) Ensemble average of number of unfolded levels in an interval of length $L$ follows $\bra n(\xi, L) \ket \approx L$. (b) Number variance $\Sigma^2(L)$ grows linearly $\approx L$ for $0 \leq L \lesssim 20$ and then oscillates with a wavelength of $\approx 200$ while saturating. Here, $\la = p_z = m = k = \mu = 1$.}
	\label{f:num-avg-and-num-variance}
	\end{figure}

Associated to the unfolded spectrum $\xi_{1}, \xi_2, \ldots \xi_N$, we have the Dirac-comb spectral density function $d(\xi) = \sum_{i =1}^N \delta(\xi - \xi_i)$.  Given a spectral window $[ \xi-L/2, \xi + L/2]$, the number of energy levels it contains is: 
	\beq
	n(\xi, L) = \int_{\xi - L/2}^{\xi + L/2} d(\xi') \; d\xi'.
	\eeq
The ensemble average $\bra n(\xi, L) \ket$ is defined as the mean value of $n(\xi, L)$ as $\xi$ ranges over an appropriate portion of the available spectrum: 
	\beq
	\bra n(\xi, L) \ket = (N-{2 n_0})^{-1} \sum_{j = n_0}^{N- n_0} n(\xi_j, L).
	\eeq
Here, $n_0$ is chosen so that $[\xi - L/2, \xi + L/2]$ always lies within the available spectrum. A more accurate approach is to define the ensemble average as $(\xi_{N-n_0} - \xi_{n_0})^{-1}$ times the integral of $n(\xi, L)$ over all values of $\xi$ between $\xi_{n_0}$ and $\xi_{N-n_0}$. However, the two definitions lead to roughly the same ensemble average as long as $L$ is not too small and the difference is hardly visible in our plots. If the unfolded spectrum has approximately unit mean spacing, we expect that $\bra n(\xi, L) \ket \approx L$. In Fig.~\ref{f:num-avg}, we plot $\bra n(\xi, L) \ket$ for $1 \leq L \leq 320$ by performing an ensemble average over the lowest $N = 20,000$ screwon levels, with $n_0 = 500$. $\bra n(\xi, L) \ket$ is seen to be $\approx L$ except for very small values of $L$, thus validating the unfolding procedure. 

To study fluctuations in the number of levels around its mean, we consider the number variance \cite{Stockmann}  
	\beq
	\Sigma^2(L) = \left< [n(\xi, L) - \left< n(\xi, L) \right>]^2 \right>,
	\eeq
where the ensemble average is performed as for $\bra n(\xi, L) \ket$. In Fig~\ref{f:num-variance}, we plot $\Sigma^2(L)$ for the lowest $N=20,000$ screwon levels of the RR model, with $\la = 1$.
For small $L \lesssim 20$, the number variance $\Sigma^2 \approx L$. For larger $L$,  $\Sigma^2$ saturates and  oscillates. 

Here we interpret the result that $\Sigma^2(L) \approx L$ for small $L$ in terms of the 2-point correlation function of the spectral density $d(\xi)$. Indeed, note that
	\beq
	\Sigma^2(L) = \left< (n(\xi, L) - \left< n(\xi, L) \right>)^2 \right> = \bra n^2 \ket + \bra n \ket^2 - 2 \bra n \bra n \ket \ket. 
	\eeq
Now using $\bra n \ket \approx L$ and $\Sigma^2(L) \approx L$, we deduce that $\bra n(\xi, L)^2 \ket = L^2 + L$. From the definition
	\beq
	\bra n(\xi , L)^2 \ket = \int_{\xi - L/2}^{\xi + L/2} d\xi_1 \int_{\xi - L/2}^{\xi + L/2} d\xi_2 \bra d(\xi_1) d(\xi_2) \ket,
	\eeq
we see that if
	\beq
	\bra d(\xi_1) d(\xi_2) \ket = 1 + \delta(\xi_1 - \xi_2)
	\eeq
then $\bra n(\xi , L)^2 \ket = L^2 + L$. This may further be interpreted as saying the connected 2-point correlation function of $d(\xi)$ is a delta function. To see this, recall that
	\beq
	\bra d(\xi_1) d(\xi_2) \ket_c = \bra d(\xi_1) d(\xi_2) \ket - \bra d(\xi_1) \ket \bra d(\xi_2) \ket.
	\eeq
Thus we see that if $\bra d(\xi_1) d(\xi_2) \ket_c = \del(\xi_1 - \xi_2)$ and $\bra d(\xi_1) \ket = 1$, then $\bra n(\xi, L)^2 \ket = L^2 + L$ and consequently $\Sigma^2(L) \approx L$. The condition $\bra d(\xi_1) \ket = 1$ is expected since the mean level density of the unfolded spectrum is 1. Of course, this linear growth of $\Sigma^2(L)$ is valid only for small $L$. 

It would be interesting to understand the subsequent saturation and oscillations in $\Sigma^2(L)$ possibly in terms of short periodic orbits of the Rajeev-Ranken model in a semiclassical approximation.

\subsection{Spectral rigidity}
\label{s:spectral-rigidity}

The spectral rigidity $\Delta_3(L, E)$ \cite{Berry} is a measure of fluctuations in the cumulative spectral distribution (staircase function) $n(E)$. It is defined as the `local average' of the mean square deviation of the best fit straight line to $n(E)$ (\ref{e:stair-case-function}) over an energy window $W = [E-L/2\bar d(E), E+L/2\bar d(E)]$:
	\beqs
	\Delta_{3}(L,E) &=& \bra \sig_*(E,L) \ket
	\quad \text{where} \quad 
	\sig_*(E,L) =  {\rm min}_{A,B} \sig(E,L,A,B) 
	\quad \text{and} \cr
	\sigma(E,L,A,B) &=& (\bar d(E) / L) \int_W \left[ n(E+\eps) -(A \eps +B) \right]^{2} {\rm d}\eps.
	\label{e:spectral-rigidity}
	\eeqs
Given a central energy $E$, the best fit is computed over an energy range $L$ in units of the local mean spacing $\bar d(E)$. We take $\bar d(E)$ to be the smoothed density of states
	\beq
	\bar d(E) = \frac{1}{2 \delta} \int_{E-\delta}^{E+\delta} d(E') {\rm d}E',
	\eeq
which is a measure of the mean spacing between levels in the $\delta$-vicinity of energy $E$. We comment on the choice of $\delta$ below. Having obtained the best fit straight line $A_* \eps + B_*$, we calculate the extremal mean square deviation $\sig_*$ for the chosen $E$ and $L$. Finally, the local average indicated by $\bra \cdots \ket$ in (\ref{e:spectral-rigidity}) is evaluated by  averaging $\sig_*(E,L)$ over a range of energy levels for fixed $L$.

\begin{figure}[h]
	\begin{center}
		\begin{subfigure}[t]{5.3cm}
		\centering
		\includegraphics[width=5.3cm]{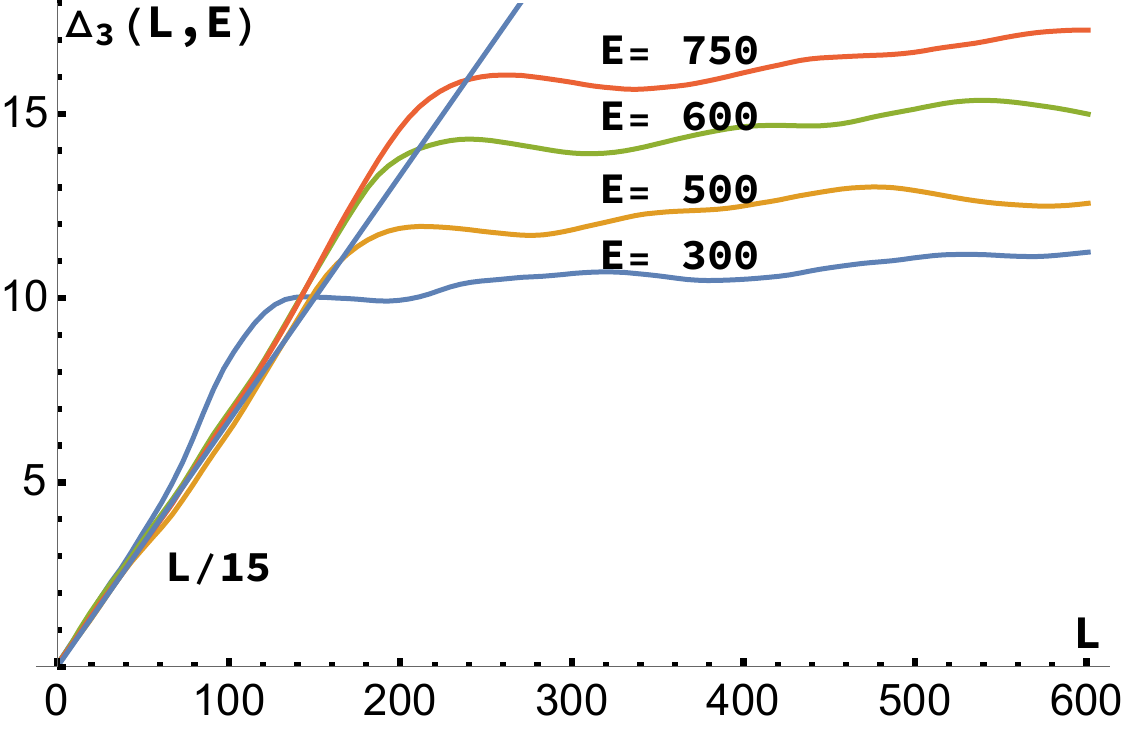}
		\caption{}
		\label{f:spectral-rigidity}
		\end{subfigure}
		\quad
		\begin{subfigure}[t]{5.7cm}
		\centering
		\includegraphics[width=5.7cm]{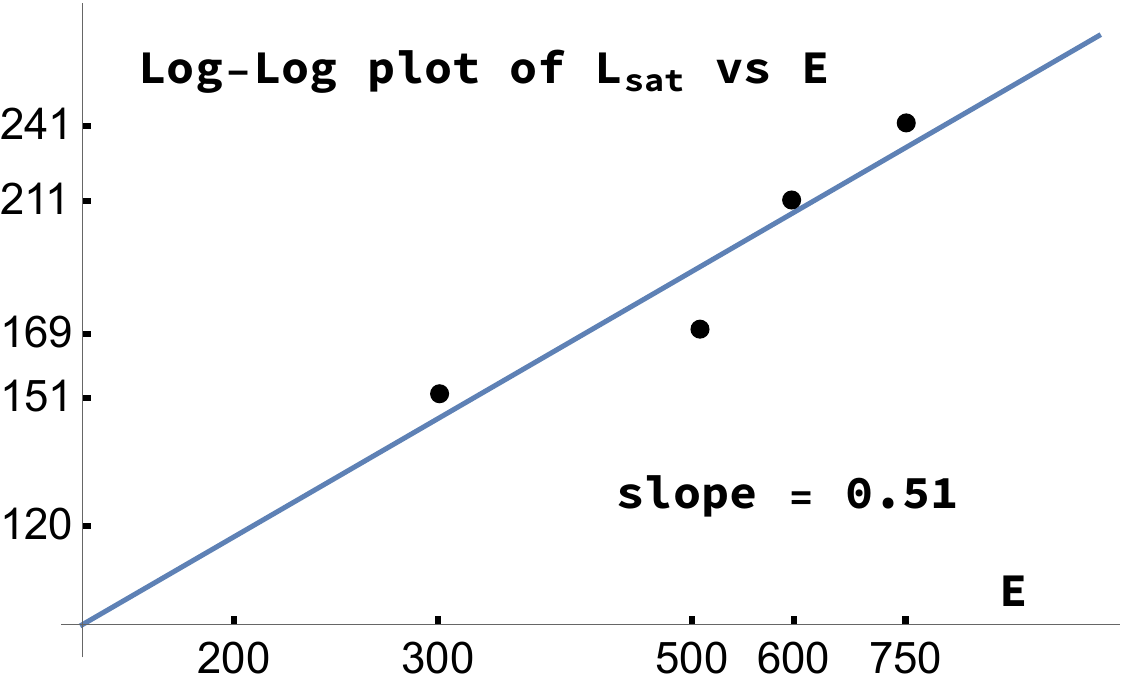}
		\caption{}
		\label{f:Lsat-vs-E}
		\end{subfigure}
		\quad
		\begin{subfigure}[t]{5.7cm}
		\centering
		\includegraphics[width=5.7cm]{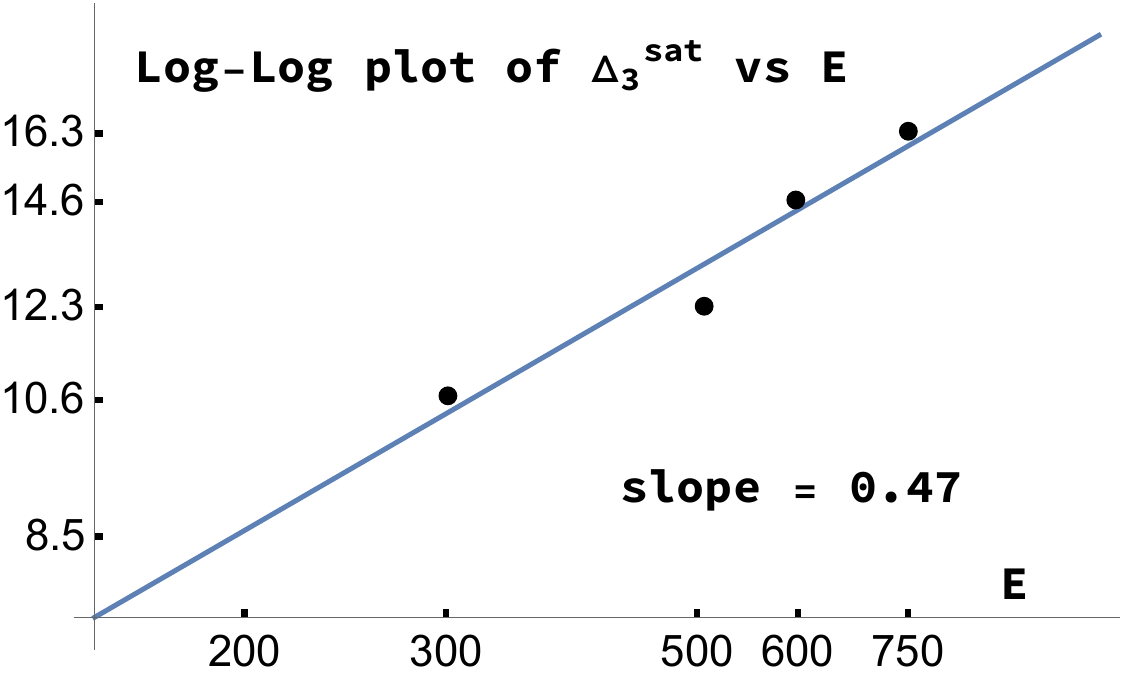}
		\caption{}
		\label{f:Delta-3-sat-vs-E}
		\end{subfigure}
		\end{center}
		\caption{\small (a) Spectral rigidity $\D_3(L,E)$ for the screwon spectrum for central energies $E = 300, 500, 600$ and $750$ (in units where $\mu = m = \hbar = 1$)  follow the universal $L/15$ law for small $L$ and then saturate around $\D_3^{\rm sat}(E)$ beyond $L= L_{\rm sat} (E)$.  Here, for definiteness $L_{\rm sat}(E)$ is defined as the largest value of $L$ for which $\D_3(L, E)$ intersects the  $L/15$ straight-line for a given $E$. Moreover, $\D_3^{\rm sat}(E)$ is defined as the mean value of $\D_3$ over the range $L_{\rm sat} \leq L \leq L_{\rm max} = 600$. (b) and (c) Log-Log plots of $L_{\rm sat}$ and $\D_3^{\rm sat}$ vs $E$ show that both display $\sqrt{E}$ behavior. In all plots we have chosen $\la = k = p_z = 1$.}
		\label{f:spec-rigidity}
	\end{figure}

For the quantum RR model, we have calculated (see Appendix \ref{a:spectral-rigidity}) the spectral rigidity $\D_3(L,E)$ for $L \leq 600 = L_{\rm max}$ for several central energies $E = 300, 500, 600, 750$ and displayed the results in Fig.~\ref{f:spec-rigidity}. In estimating the local mean spacing $\bar{d}(E)$, we choose $\delta$ to accommodate about 80 levels centered at $E$. The value of $\bar{d}(E)$ is largely insensitive to the choice of $\delta$. On the other hand, the local average is performed over an energy interval $[E - 20, E + 20]$. The results are insensitive to small changes in the energy range for the local average. In fact, essentially the same results are obtained if the local average is performed over an ensemble of about 2000 levels centered at $E$. Although $\D_3(L,E)$ generally depends on the central energy $E$, for small $L$ it approaches the universal linear shape $\D_3(L,E) \to L/15$ as expected from Berry's arguments \cite{Berry} for integrable systems based on the semi-classical trace formula applied to long-periodic orbits. Beyond a critical value of $L = L_{\rm sat}$, the spectral rigidity oscillates around a saturation value $\D_3^{\rm sat}(E)$. Both $L_{\rm sat}$ and $\D_3^{\rm sat}$  increase when the central energy $E$ is significantly augmented [due to fluctuations, this monotonicity can fail for energies that are not widely separated]. In fact, as shown in Figs.~\ref{f:Lsat-vs-E} and \ref{f:Delta-3-sat-vs-E}, we find that both $L_{\rm sat}$ and $\D^{\rm sat}_3$ are approximately proportional to the square-root of energy. The latter power law is expected from the work of Casati, Chirikov and Guarneri \cite{ Stockmann, C-C-G}.


\section{Dispersion relation at strong coupling}
\label{s:dispersion-relation-screwon}

In \cite{G-V-3}, we had conjectured a $(\la k)^{2/3}$ power law dispersion relation for the energies of highly excited screwons at strong coupling based on a numerical inversion of the WKB quantization condition. Here, we generalize this conjecture, make it more precise and provide additional evidence for it going beyond the WKB approximation. 

Recall that the quartic potential $U = \al r^2 + \beta r^4$ in the Hamiltonian (\ref{e:Hamiltonian-quadratic-quartic-Cartesian}) involves the coefficients $\al$ and $\beta$ (\ref{e:alpha-and-beta}). At strong coupling, by which we mean 
	\beq
	\la \gg 2\sqrt{\mu}/m \quad \text{and} \quad \la \gg \mu k/p_z,
	\label{e:strong-coupling-conditions}
	\eeq
we may neglect the $k^2/2$ term in $\alpha$, so that $U$ depends on $\la$ and $k$ only through their product. In fact, in this limit, the shifted radial Hamiltonian $H-k^2m^2/2 - p_z^2/2 \mu$  (\ref{e:radial-equation-dimensionful-varrho}) also depends on $\la$ and $k$ only through their product. Thus, the shifted energy eigenvalues $E- k^2 m^2/2 - p_z^2/2\mu$ determined by the radial equation \small
	\beq
	-\frac{\hbar^2}{2 \mu} \left(  \frac{{\rm d^2}}{{\rm d}r^2}  + \frac{1}{r} \frac{{\rm d}}{{\rm d}r} - \frac{l^2}{r^2} \right) \rho(r)+ \frac{\la^2 k^2}{8 \mu} \left(\left( m^2 - \frac{4 p_z}{\la k} \right) r^2 + r^4 \right)\rho + \frac{\hbar l \la k m}{2 \mu}  \rho \approx \left( E- \frac{k^2 m^2}{2} - \frac{p_z^2}{2 \mu}  \right) \rho
	\label{e:rad-eqn-str-coupl-naive}
	\eeq \normalsize
must be a function of $\la k$. More precisely, on dimensional grounds,  
	\beq
	E_{n, l} \approx \frac{p_z^2}{2\mu} + \frac{k^2 m^2}{2} + \frac{\hbar^2}{\mu m^2} \eps_{n, l}\left(\tl \sig, \frac{m p_z}{\hbar} \right) \quad \text{for large} \quad \la \quad \text{satisfying} \quad (\ref{e:strong-coupling-conditions}). 
	\eeq
Here, $\hbar^2 \eps_{n, l}/\mu m^2$ are the eigenvalues of $H- k^2 m^2/2 - p_z^2/2\mu$ for large $\la$ (satisfying (\ref{e:strong-coupling-conditions})) and $\eps_{n, l}$ is a function of the dimensionless variables $\tl \sig = \la k m^3/\hbar$ and $m p_z/\hbar$. In the limit (\ref{e:strong-coupling-conditions}) the other independent dimensionless combination $\sqrt{\mu}/m \la$ does not enter as it tends to $0$. Moreover, if
	\beq
	\frac{4p_z}{m^2 \la} \ll k \ll \frac{\la p_z}{\mu},
	\label{e:condition-on-k}
	\eeq
then $\eps_{n, l}$ becomes independent of $p_z$ as the $4p_z/\la k$ term can be ignored relative to $m^2$ in (\ref{e:rad-eqn-str-coupl-naive}).

\begin{figure}[h]
	\begin{center}
		\begin{subfigure}[t]{7cm}
		\centering
		\includegraphics[width=7cm]{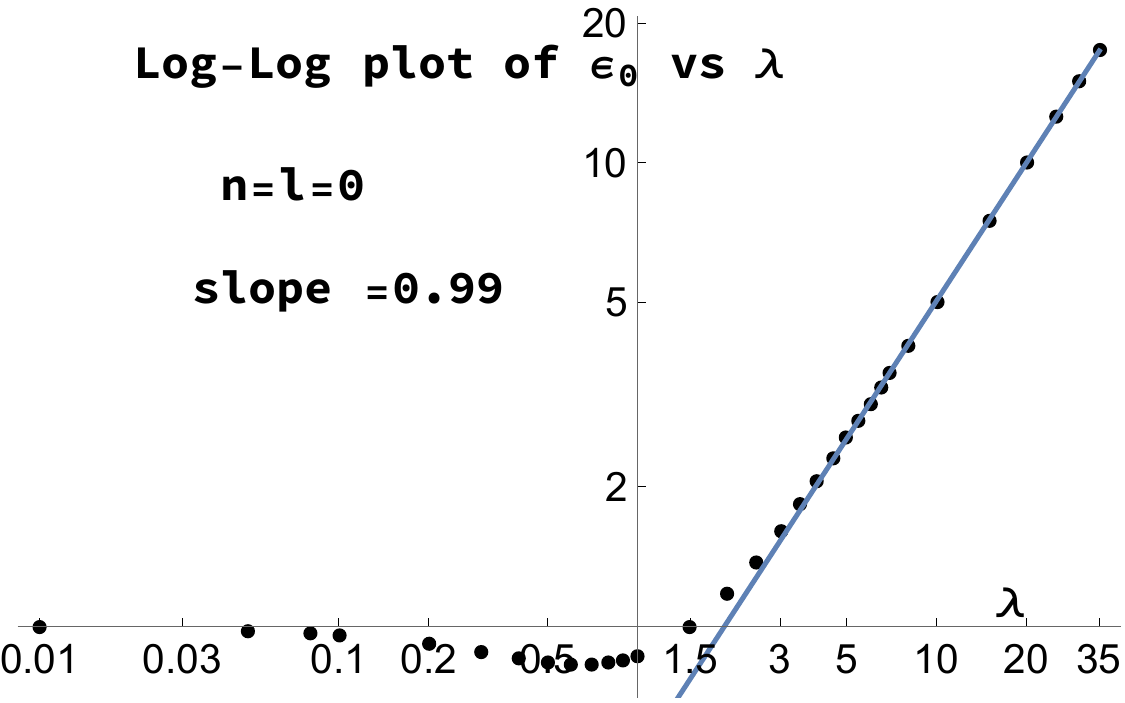}
		\caption{}
		\label{f:LogE-vs-Log-Lambda-small-n}
		\end{subfigure}
		\qquad
		\begin{subfigure}[t]{7cm}
		\centering
		\includegraphics[width=7cm]{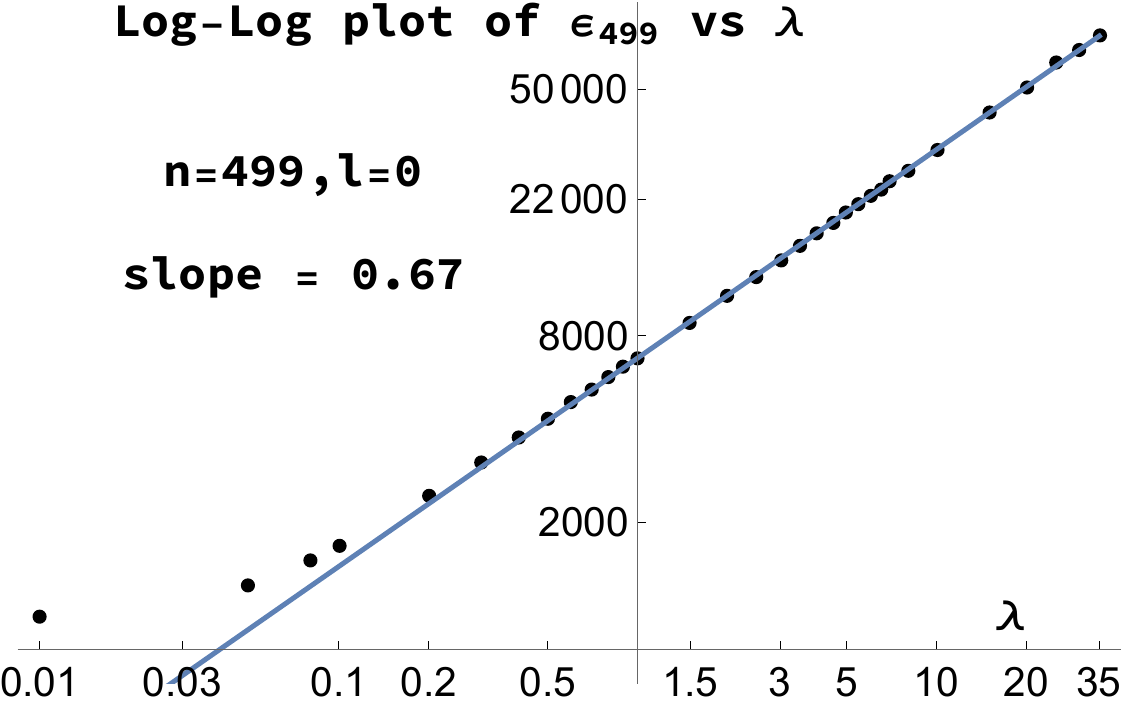}
		\caption{}
		\label{f:LogE-vs-Log-lambda}
		\end{subfigure}		
		\qquad
		\begin{subfigure}[t]{7cm}
		\centering
		\includegraphics[width=7cm]{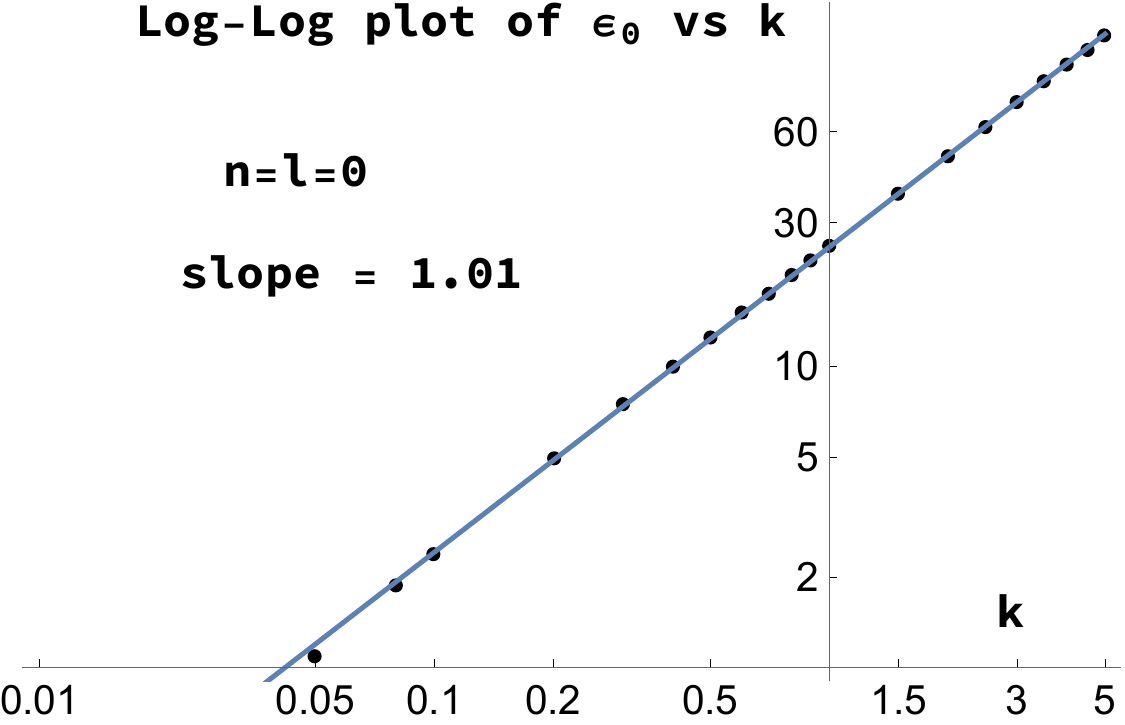}
		\caption{}
		\label{f:LogE-vs-Log-k}
		\end{subfigure}
		\qquad
		\begin{subfigure}[t]{7cm}
		\centering
		\includegraphics[width=7cm]{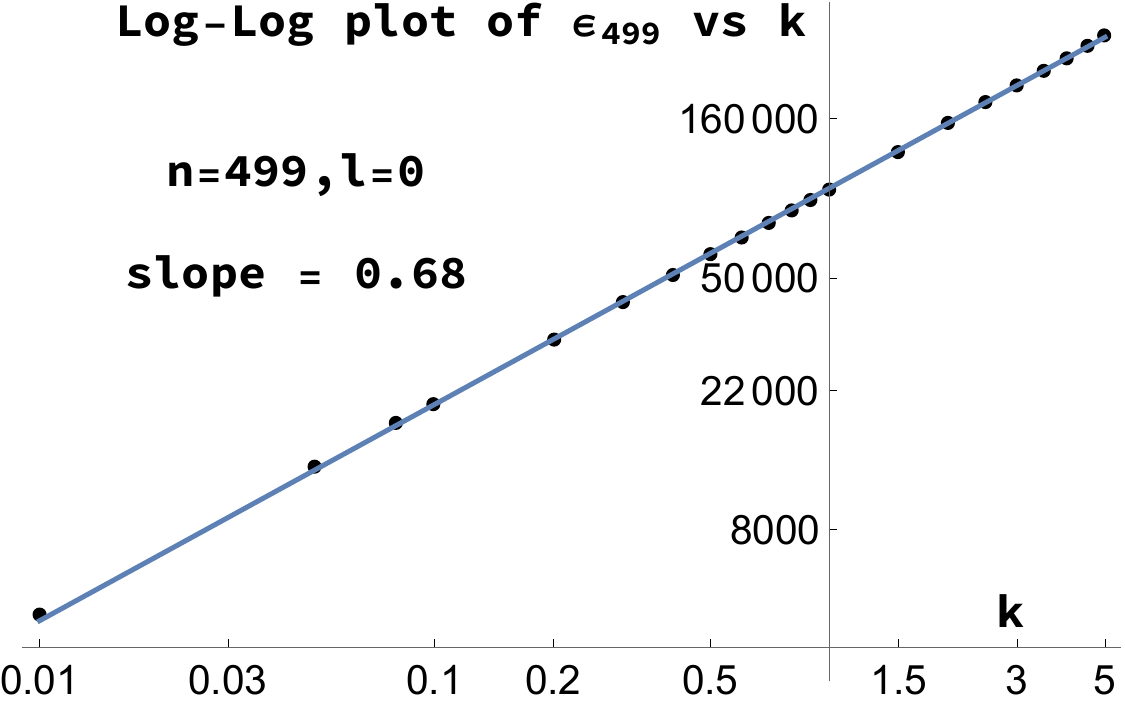}
		\caption{}
		\label{f:LogE-vs-Log-k-large-n}
		\end{subfigure}
	\end{center}
		\caption{\footnotesize $\log \eps_{n,l}$ vs $\log \la$ for (a) $n= 0$ and (b) n = 499 with $l= 0$ when $k =1$. For sufficiently large $\la$, the slope of the fitted straight line decreases from $1$ to $2/3$ as we progress from the ground state to excited states.  $\log \eps_{n,l}$ vs $\log k$ for (c) $n = 0$ and (d) $n =499$ with $l =0$ for $\la = 50$. For $k$ satisfying (\ref{e:condition-on-k}) and sufficiently large $\la$, we find that $\eps_{n,l} \propto k^{\eta}$ with the power decreasing from $ \eta \approx 1$ to $\eta \approx 2/3$ as we progress from the ground to excited states. In all cases, we take $ p_z = m = \mu = \hbar =1$.}
		\label{f:dispersion-relations-Log-Log-plots}
	\end{figure}

From numerical diagonalization of $H$, we find for fixed $l$ and $p_z$ at strong coupling and moderate $k$ satisfying (\ref{e:strong-coupling-conditions}) and (\ref{e:condition-on-k}) that $\eps_{n, l}$ depends on both $\la$ and $k$ via power laws (see Fig.~\ref{f:dispersion-relations-Log-Log-plots}). For the ground state ($n = l = 0$) the powers of $\la$ and $k$ are approximately equal to $1$. The powers decrease with increasing level number $n$ (holding $l =0$ fixed) and rapidly approach the common value $\eta =2/3$ for sufficiently excited states ($n \gtrsim 100$). It would be interesting to analytically understand the emergence of this $2/3$ power law for highly excited screwons at strong coupling.

We have not investigated the $l$-dependence of the exponent in detail. However, we numerically verified that for $n \gg |l|$, $\eps_{n, l} \propto (\la k)^{\eta}$  where $\eta \approx 2/3$ is independent of $l$. For instance, this holds if $n \gtrsim 200$ and $|l| \leq 20$, for $0.08 \leq k \leq 5$ holding $\la = 50$ fixed as well as for $4 \lesssim \la \lesssim 35$ holding $k = 1$ fixed. However, if $|l|$ is comparable to or greater than $n$, then $\eta$ depends on $l$. For example, if $n =0$, then $\eta \approx 1, 0.4,0.9$ for $l= 0,-20,20$. 

\section{Discussion}
\label{s:discussion}

In this paper, we have obtained results on the spectral statistics and energy-wavenumber dispersion relation of screwons in the quantum Rajeev-Ranken model. We now discuss possible physical implications and outstanding questions arising from this work. 

The RR model is a mechanical reduction describing screw-type nonlinear waves in a 1+1-dimensional nilpotent scalar field theory \cite{R-R}. This parent scalar field theory has a perturbative Landau pole (where the coupling in the perturbative approximation blows up) and is strongly coupled in the ultraviolet. Thus, we do not know the degrees of freedom that govern its high-energy behavior. Our results show that the spectrum of quantized screwons extends to arbitrarily high energies in the RR model. This gives us hope that screwons could play a role in the parent scalar field theory at asymptotically high energies. What is more, the fractional power law dispersion relation that we have found could point to novel behavior of these quantized screwons unlike the free scalar particle-like behavior (see Eqn.~(\ref{e:EOM-scalar-field-theory})) of weakly coupled excitations at low energies. Our study of spectral statistics has revealed quantum signatures of integrability in the RR model.  This motivates us to look for integrable structures in the scalar field theory, which could be helpful in handling the regularization and renormalization of the corresponding quantum field theory. Although the original physical applications of the RR model arise by virtue of its being a subsector of a larger scalar field theory, other physical connections are also possible. For instance, we have shown in  \cite{G-V-1} that the classical equations, Hamiltonian formulation and Lax pairs of the RR model are  structurally similar to those of the Neumann model \cite{B-B-T}, which describes the motion of a particle on a sphere subject to harmonic forces. The equations of the RR model are also similar to the Kirchhoff equations, which describe the evolution of the momentum and angular momentum of a rigid body moving in an incompressible, inviscid potential flow. In fact, the RR and Kirchhoff equations are both Euler equations for a centrally extended and nonextended Euclidean algebra  \cite{G-V-2}. Moreover, the RR model can also be re-interpreted as governing the motion of a nonrelativistic charged particle in a certain axisymmetric electromagnetic field \cite{R-R, G-V-3}. Now, our interpretation of the RR model as a novel anharmonic oscillator has allowed us to quantize it canonically without having to deal with representations of an unfamiliar nilpotent or Euclidean algebra. Furthermore, our optimized numerical scheme has enabled us to accurately compute a large portion of the spectrum of the RR model. These developments should facilitate the quantization and solution of the above related models. Moreover, our results on spectral correlations and dispersion relation should have interesting physical implications and interpretations in these other approaches where available \cite{Ba-T, Be-T-1,Be-T-2}.  
 
Returning to the specific results on the quantum RR model, there are some interesting directions for future research. Although we have recovered the expected universal behaviour of number variance and spectral rigidity at small $L$, they both display system-dependent saturation and oscillations for larger $L$. We would like to understand this nonuniversal behaviour using Gutzwiller's trace formula and Berry's semi-classical asymptotic theory \cite{Berry}. For this purpose, we intend to use the exact solutions of the classical RR model \cite{R-R, G-V-1, G-V-2} to identify and classify the shortest periodic orbits. It would also be interesting to understand the power law behaviour in the saturation of spectral rigidity and the asymptotic behaviour of the cumulative level distribution function semiclassically. In another direction, we would like to understand analytically the common power law behaviour (in both $k$ and $\lambda$) of the dispersion relation for screwons at strong coupling. This dispersion relation should shed light on the ultraviolet behaviour of screwons in the field theory.

\vspace{1cm}

{\fl \bf Acknowledgements:} We would like to thank M. V. Berry,  M.S. Santhanam and H. Senapati for useful discussions and references. In addition, we thank an anonymous referee for helpful comments. This work was supported in part by the Infosys Foundation and grants (MTR/2018/000734, CRG/2018/002040) from the Science and Engineering Research Board, Govt. of India.

\appendix

\section{Evaluating the spectral rigidity}
\label{a:spectral-rigidity}

The integrals appearing in the evaluation of spectral rigidity (\ref{e:spectral-rigidity}) are in fact finite sums. Writing them as such speeds up numerical calculations. Let us denote the limits of integration by $E_{\rm min, \rm max} = E \mp (L/2 \bar d(E))$. Then $(L/\bar{d}(E)) \sig= I_{1} + I_{2} + I_{3}$, where
	\beq
	I_1 = \int_{E_{\rm min}}^{E_{\rm max}} (A \eps + B)^2 d\eps = \left[ \frac{A^2 \eps^3}{3} + A B \eps^2 + B^2 \eps \right]_{E_{\rm min}}^{E_{\rm max}}.
	\eeq
Suppose $n_0$ is the number of levels with energy $\leq E_{\rm min}$ and $N$ is the number of levels between $E_{\rm min}$ and $E_{\rm max}$, then 
	\beq
	I_2 = \int_{E_{\rm min}}^{E_{\rm max}} n(E+ \eps)^2 d\eps = n_0^2 (E_{n_0 +1} - E_{\rm min} ) + (n_0 + N)^2 (E_{\rm max} - E_{n_0 + N}) + \sum_{k =1}^{N-1} (n_0 + k)^2 S_{n_0 + k}.
	\eeq
Here, $S_{n_0 + k} = E_{n_0+k+1} - E_{n_0 +k}$ is the level spacing. Similarly, we have 
	\small
	\beqs
	\frac{I_3}{2} &=& \int_{E_{\rm min}}^{E_{\rm max}} n(E+ \eps)(A \eps + B) d\eps = B\left[ n_0 (E_{n_0 +1} - E_{\rm min} ) + (n_0 + N) (E_{\rm max} - E_{n_0 + N}) + \sum_{k =1}^{N-1} (n_0 + k) S_{n_0 + k} \right] \cr
	&&+ \frac{A}{2} \left[n_0 (E_{n_0 +1}^2 - E_{\rm min}^2 ) + (n_0 + N) (E_{\rm max}^2 - E_{n_0 + N}^2) + \sum_{k =1}^{N-1} (n_0 + k) E_{n_0 +k+1}^2 - E_{n_0 + k}^2 \right]. \qquad 
	\eeqs
	\normalsize



\end{document}